\documentclass[%
10pt,
showkeys,
onecolumn,
amsmath,amssymb,
aps,
prb,
]{revtex4-2}

\usepackage[utf8]{inputenc} 
\usepackage[T1]{fontenc}    
\usepackage{url}            
\usepackage{booktabs}       
\usepackage{hyperref}       
\usepackage{graphicx}
\usepackage{float}
\usepackage{svg}
\usepackage[english]{babel}
\usepackage{pstricks}

\renewcommand{\textsc}[1]{{\small \uppercase{#1}}}

\usepackage{tabularx} 

\usepackage{mathtools, amsfonts, bbold} 
\usepackage{array} 
\usepackage{textcomp} 
\usepackage{siunitx} 
\usepackage{physics} 
\usepackage{amsthm} 
\usepackage{xfrac} 
\usepackage{dsfont} 

\renewcommand{\vector}[1]{\vec{#1}}

\newcommand{\partwrt}[1]{\frac{\partial}{\partial{#1}}}
\newcommand{\partofwrt}[2]{\frac{\partial{#1}}{\partial{#2}}}
\newcommand{\divwrt}[2]{\nabla_{#1} \cdot {#2}}
\newcommand{\posVector}[3]{
    \begin{bmatrix}
        #1 \\
        #2 \\
        #3
    \end{bmatrix}
    }

\newcommand{\R}{\mathbb{R}}

\begin{document}

\title{Reframing the coupling force of adaptive resolution simulation in terms of the Liouville-type hierarchy for open systems}

\author{Julian F. \surname{Hille}}
\email{julianhille@zedat.fu-berlin.de}
\author{Rupert \surname{Klein}}
\email{rupert.klein@math.fu-berlin.de}

\affiliation{Freie Universität Berlin, Institute of Mathematics, Arnimallee 6, 14195 Berlin, Germany}

\date{27.07.2026}

\begin{abstract}
In this work, we present a novel perspective on the coupling force employed to compensate the interface artifacts prevalent in adaptive resolution simulations (AdResS) of open many-particle systems. We show that a substantial part of this ``thermodynamic force'' can be framed in terms of the theoretical model of the Liouville-type hierarchy for open systems. The correspondence is made explicit for the case of a simple atomistic fluid, for which a one-dimensional integral expression is derived. This enables the analysis for dependencies of the thermodynamic force on important simulation parameters, which is taken to inspire both simplifications for the numerical calculation of the thermodynamic force and new criteria for its validation that are adequate to the interfacial nature of the problem. The theoretical claims are then verified in a simulation study of the atomistic supercritical Lennard-Jones fluid at different thermodynamic states.  
\end{abstract}

\keywords{Adaptive resolution simulation; Statistical mechanics; Molecular dynamics; Open molecular systems}

\maketitle

\section{Introduction}
The adaptive resolution simulation scheme (AdResS) comprises a molecular dynamics algorithm to simulate an open system exchanging particles and energy with its environment\cite{praprotnik_adaptive_2005, praprotnik_multiscale_2008, delle_site_molecular_2017}. 

In its current version, this is accomplished by partitioning the simulation domain into a region within which the particles interact via pair-wise potentials and a region containing only non-interacting tracer particles (\textsc{TR})\cite{delle_site_molecular_2019}. The interactions are switched on and off accordingly for each particle crossing one of the abrupt interfaces in between these regions. The inner interacting, atomistic (\textsc{AT}), region has been shown to exhibit grand-canonical statistics if properly equilibrated\cite{wang_grand-canonical-like_2013, gholami_thermodynamic_2021, gholami_relation_2022}. That is, if the artifacts that result from the switching of interactions are being compensated within the narrow transition (\textsc{$\Delta$}) regions in between the boundaries of the \textsc{AT} region and the abrupt interfaces. 

This is achieved through the application of an external coupling force, the ``thermodynamic force'', together with a stochastic thermostat and a force-capping scheme\cite{krekeler_adaptive_2018}. The well-established iterative procedure to find the thermodynamic force involves itself multiple shorter AdResS simulations\cite{poblete_coupling_2010, fritsch_adaptive_2012, wang_grand-canonical-like_2013}. The force then has to be verified \textit{a posteriori} against a full-atomistic reference simulation in an extensive validation run\cite{gholami_thermodynamic_2021, gholami_relation_2022}. 

The methodological progress of AdResS has been accompanied by the development of the Liouville-type hierarchy for open systems\cite{delle_site_liouville-type_2020, klein_derivation_2022}. Mapping of AdResS to this model provides a clear statistical mechanical understanding of the AdResS \textsc{AT} region as an open system and of the thermostatted \textsc{$\Delta$} and \textsc{TR} regions with the applied thermodynamic force as a particle and energy reservoir. Such theoretical treatment was inspired by the Bergmann-Lebowitz model of open systems, to which AdResS had previously been mapped qualitatively\cite{bergmann_new_1955, lebowitz_irreversible_1957, agarwal_molecular_2015}.

AdResS has been shown capable of simulating non-equilibrium phenomena by employing thermodynamically distinct reservoirs at opposing boundaries of the \textsc{AT} region and applying the according thermodynamic forces\cite{ebrahimi_viand_theory_2020, klein_nonequilibrium_2021}. The inherent multi-scale nature of the method has inspired composite schemes that couple classical molecular dynamics with quantum mechanics\cite{poma_classical_2010, panahian_jand_nuclear_2022, panahian_jand_physical_2024} or with fluid dynamics, where the reservoir states and the respective thermodynamic forces are time-dependent\cite{gholami_simulation_2022}. In that regard, AdResS comprises one among several particle-continuum coupling schemes that have emerged in recent years\cite{donev_hybrid_2010, delgado-buscalioni_open_2015, delle_site_particlecontinuum_2020, smith_multiscale_2024, lah_open-boundary_2025, djurdjevac_hybrid_2025}.  

Building upon the mapping of AdResS to the Liouville-type hierarchy, we show here that there are clear requirements on the conditions at the boundaries of the \textsc{AT} region that have immediate implications on the thermodynamic force. Starting from the boundary terms appearing in the hierarchy, we derive an integral expression in light of which the thermodynamic force can be reinterpreted and from which we can read off dependencies on important simulation parameters. This then motivates simplifications for the iterative procedure and new validation criteria that are more specific to the interfacial nature of the problem. We verify the validity of our claims in a simulation study of the Lennard-Jones fluid at three different thermodynamic states.
\section{Simulation method and theory for open systems}\label{sec:simulation_method_and_theory}
In this work we will employ the common AdResS setup with rectangular simulation domain and periodic boundary conditions in all directions\cite{delle_site_molecular_2019}. 

Before partitioning the domain into the \textsc{AT}, \textsc{$\Delta$} and \textsc{TR} regions, the system is equilibrated to the desired thermodynamic state with standard procedures\cite{braun_best_2019}, employing an atomistic radial potential $V(r)$ for all particles throughout the simulation domain, where $r$ is the distance between any two particles $i$ and $j$. 

\begin{figure}[h]
    \centering
    \input{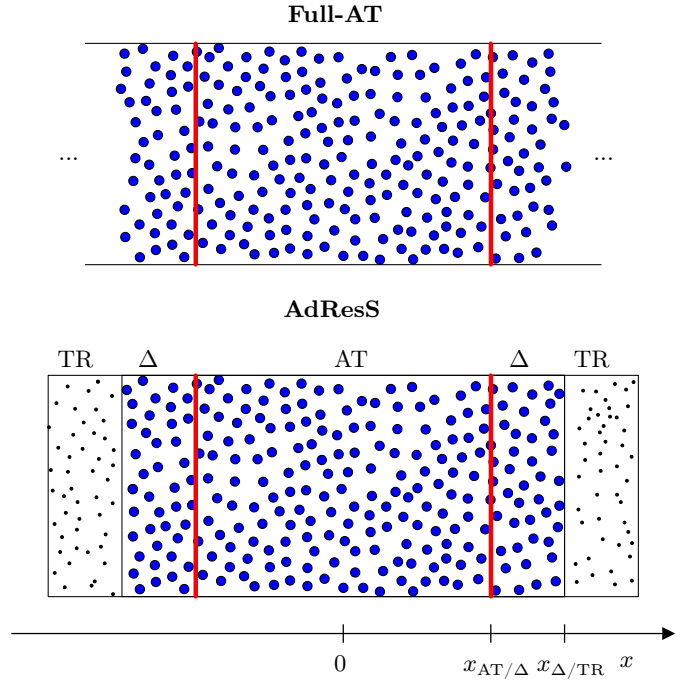}
    \vspace{-1.2cm} 
    \caption{Different setups for the simulation of an open molecular system. Full-\textsc{AT}: Reference setup of the open system represented by a small subsystem in a larger fully-atomistic simulation. AdResS: Adaptive resolution simulation setup of the open system by enclosing a small subsystem in between two \textsc{AT/$\Delta$} boundaries; Within $\textrm{\textsc{AT}} \cup \textrm{\textsc{$\Delta$}}$, the particles interact with each other via the same pair-wise potential as in the full-\textsc{AT} reference; In the \textsc{TR} region, the particles have no interactions, giving rise to artifacts of the \textsc{$\Delta$/TR} interfaces that need to be compensated; The $x$-axis is normal to the \textsc{AT/$\Delta$} boundaries and \textsc{$\Delta$/TR} interfaces. The boundaries of the open system are marked in red in both graphics.}
    \label{fig:illustration_adress}
\end{figure}

In the equilibrated system, two planar and abrupt \textsc{$\Delta$/TR} interfaces parallel to the $yz$-plane are then introduced by replacing $V(r)$ with a new pair-wise potential 
\begin{align}
    U(r, x_i, x_j) = \begin{cases}
            V(r) &\textrm{, for } x_i, x_j \in [-x_{\textrm{$\Delta$/TR}}, x_{\textrm{$\Delta$/TR}}] \\
            0 &\textrm{, else } \\
            \end{cases}\textrm{,}
\end{align}
where $x_i$ and $x_j$ are the positions of $i$ and $j$ in $x$-direction. This is illustrated in the lower graph of Fig. \ref{fig:illustration_adress}. For convenience, we place the center of the simulation box onto the origin such that the right-hand \textsc{$\Delta$/TR} interface is at $x_{\textrm{$\Delta$/TR}}$ and the left-hand one is at $-x_{\textrm{$\Delta$/TR}}$. The planar geometry of the \textsc{AT/$\Delta$} boundaries positioned at $x_{\textrm{AT/$\Delta$}}$ and $-x_{\textrm{AT/$\Delta$}}$ is used here to simplify our theoretical derivations, though AdResS could also be used with other geometries such as a spherically symmetric \textsc{AT} region\cite{krekeler_adaptive_2018, shadrack_jabes_ionic_2018, panahian_jand_nuclear_2022}. 

The simulation run with the new potential $U(r, x_i, x_j)$ will suffer from numerical artifacts of the \textsc{$\Delta$/TR} interfaces. These hence need to be mitigated within the \textsc{$\Delta$} regions such that the \textsc{AT} region is subject to a physically sound environment and exhibits the properties of an open system\cite{delle_site_molecular_2019}. 

A common verification procedure is to compare physical observables such as the particle number fluctuations and radial distribution functions measured in the \textsc{AT} region of AdResS with the same observables obtained in an equivalent subregion of a full-atomistic (full-\textsc{AT}) reference simulation, as depicted in the upper graph of Fig. \ref{fig:illustration_adress}\cite{ciccotti_physics_2019, gholami_thermodynamic_2021, gholami_relation_2022}.

\subsection{The AdResS interface correction}\label{sec:interface_corrections}
The correction of the interface artifacts can be achieved by a Langevin thermostat, a force capping scheme and an external one-body thermodynamic force\cite{delle_site_molecular_2019}.

The role of the thermostat is to dissipate the momentum leaking into the simulation box due to tracer particles entering arbitrarily close to particles in the \textsc{$\Delta$} regions, which gives rise to enormous repulsive forces\cite{delle_site_molecular_2017}. To prevent these forces from breaking the simulation before the thermostat can dampen them, the maximum possible force is capped at a suitable value\cite{kreis_advantages_2015, delle_site_molecular_2019}. This capping may either be performed on the particle-wise forces or on the pair-wise forces, which amounts to replacing the short-range repulsive part of the radial potential $V(r)$ with a linear extrapolation beneath a characteristic radius $r_{\textrm{cap}}$\cite{krekeler_adaptive_2018, delle_site_molecular_2019}. As the AdResS procedure is built on a local conception of resolution, it is customary to use a truncated potential with a cut-off radius $r_{\textrm{cut}}$\cite{heidari_accurate_2016, delle_site_molecular_2019}. We can thus express the general form of the radial interparticle potential employed in the interacting region of AdResS simulations with pair-wise capping scheme as
\begin{align}\label{eq:V_cut_cap}
    V(r) = \begin{cases}
            v(r_{\textrm{cap}}) - \partofwrt{v}{r}\big|_{r_\textrm{cap}} r_{\textrm{cap}} + \partofwrt{v}{r}\big|_{r_\textrm{cap}} r &\textrm{, for } r_{\textrm{cap}} \geq r \\
            v(r) &\textrm{, for } r_{\textrm{cap}} < r \leq r_{\textrm{cut}} \\
            0 &\textrm{, for } r_{\textrm{cut}} < r \\
            \end{cases}\textrm{,}
\end{align}
where $v(r)$ is a suitable pair-wise potential for molecular dynamics simulations such as the truncated and shifted Lennard-Jones potential. More sophisticated algorithms for the particle insertion have been developed that do not require capping of the forces or even thermostatting but are computationally more complex and/or expensive than the one considered here\cite{peters_simulation_2016, thaler_back-mapping_2020}.

The thermodynamic force then needs to compensate for (1) the residual effects of the \textsc{$\Delta$/TR} interfaces that have not been compensated fully by the thermostat, (2) the thermodynamic differences between the \textsc{AT} and \textsc{TR} regions and (3) the missing forces that particles residing in the \textsc{$\Delta$} regions would be subject to if there were no \textsc{$\Delta$/TR} interfaces and the simulation would be fully atomistic. It can be calculated via an iterative procedure according to
\begin{align}\label{eq:thermo_force_iterations}
    F_{\textrm{th}}^{k + 1}(x) &= F_{\textrm{th}}^{k}(x) - c \nabla \rho^{k}(x) \textrm{,}
\end{align}
where each iteration $k$ comprises a short AdResS simulation with applied thermodynamic force $F_{\textrm{th}}^{k}(x)$. Throughout any iteration, the instantaneous density profile is measured frequently by counting the numbers of particles residing within narrow bins along the $x$-direction and dividing by the respective bin volume. The density profile $\rho^{k}(x)$ is then obtained by averaging over the profiles of all these snapshots after sufficient time has elapsed. Its negative gradient weighted by a convergence prefactor $c$ is then applied as a correction to the thermodynamic force and used for the next iteration\cite{krekeler_adaptive_2018}. The thermodynamic force is considered converged when the AdResS simulation produces a density profile flat to within a prescribed tolerance. 

It is clear that the \textsc{AT} region can only be a physically sound open system if the artifacts caused by the \textsc{$\Delta$/TR} interfaces have been compensated within the \textsc{$\Delta$} regions such that the particles arriving at the \textsc{AT/$\Delta$} boundaries behave sufficiently close to fully-atomistic. What exact criteria have to be met by the particles at the \textsc{AT/$\Delta$} boundaries to adhere to that requirement can be understood by considering the Liouville-type hierarchy for open systems\cite{delle_site_liouville-type_2020, klein_derivation_2022}.

\subsection{Liouville-Type Hierarchy for Open Systems}
The time evolution of the phase space probability density $F_N$ of a universe with spatial domain $U$ containing $N$ particles is governed by the well-known Liouville equation. An analogous Liouville-type equation for the probability density $f_n$ of $n$ particles residing within an open subsystem $\Omega \subset U$ can be obtained by marginalizing the $N$-particle Liouville equation with respect to the degrees of freedom of all $N - n$ particles outside of the open system, i.e. in $\Omega_c = U \setminus \Omega$\cite{delle_site_liouville-type_2020}. The marginalized equation then reads
\begin{align}\label{eq:liouvilleTypeHierarchy}
    \partwrt{t}{f_n} + \sum_{i = 1}^{n}\left(\divwrt{\vector{q}_i}{(\vector{v}_i f_n)} + \divwrt{\vector{p}_i}{(\vector{F}_i f_n)}\right) &= \Psi_n + \Phi_n^{n + 1} \textrm{,}
\end{align}
where $\vector{q}_i$, $\vector{p}_i$, $\vector{v}_i$ are the position, velocity and momentum vectors of the $i$-th particle in $\Omega$, respectively, and
\begin{align}\label{eq:innerTotalForce}
    \vector{F}_i &= -\sum_{j > i}^{n} \nabla_{\vector{q}_i} V(\vector{q}_i - \vector{q}_j)
\end{align}
is the force acting on the $i$-th particle resulting from the two-body interactions $V(\vector{q}_i - \vector{q}_j)$ with all $n - 1$ other particles $j$ within $\Omega$.  

As the number of particles changes dynamically, probability density is transported across the boundary of the open system. This gives rise to source terms that couple each Liouville-type equation with the environment and with the corresponding equations of different $n$ in a hierarchical ordering. The first such term on the r.h.s. of Eq. (\ref{eq:liouvilleTypeHierarchy}) is given as
\begin{align}\label{eq:psi}
    \Psi_n &= -\sum_{i = 1}^{n}\divwrt{\vector{p}_i}{\left(\vector{F}_{\textrm{av}}(\vector{q}_i; \partial \Omega) f_n\right)} \textrm{,}
\end{align}
where 
\begin{align}\label{eq:avForce} 
    \vector{F}_{\textrm{av}}(\vector{q}_i; \partial \Omega) &= - \int_{\Omega_{\textrm{c}}} \dd{\vector{q}_k} \int_{\R^3} \dd{\vector{p}_k} \nabla_{\vector{q}_i} V(\vector{q}_i - \vector{q}_k) f_{k|i}^0
\end{align}
is an average force acting on a particle $i$ within the open system resulting from the pairwise interactions with the particles on the outside $\Omega_{\textrm{c}}$ and through the open system's boundary $\partial \Omega$. The function $f_{k|i}^0$ appearing in the integrand is the conditional equilibrium probability density of a representative reservoir particle $k$ being at position $\vector{q}_k$ with momentum $\vector{p}_k$ given the state of the inner particle $i$. As we will show in the next section, for the simple atomistic fluids considered here, $f_{k|i}^0$ is independent of $\vector{p}_i$ and thus so is the average force $\vector{F}_{\textrm{av}}$.

The second source term of Eq. (\ref{eq:liouvilleTypeHierarchy}) facilitates the coupling in between neighboring equations of the hierarchy through the exchange of particles between the open system and the reservoir and is given as 
\begin{align}\label{eq:phi}
    \Phi_n^{n + 1} &= (n + 1) \int_{\partial \Omega} \dd{\sigma_k} \int_{(\vector{v}_k \cdot \hat{n}) > 0} \dd{\vector{p}_k} (\vector{v}_k \cdot \hat{n}) \left(f_{n + 1} - f_{n + 1}^{\textrm{in}}\right) \textrm{,}
\end{align}
where the outer integration is performed over the open system's boundary with surface normal vector $\hat{n}$ and surface element $\dd{\sigma_k}$ and the inner integration is over all velocities $\vector{v}_k$ pointing outward of the open system. Following the modeling assumptions of ref. \cite{delle_site_liouville-type_2020}, the probability density $f_{n + 1}^{\textrm{in}}$ representing the influx of particles through the boundary can be written as 
\begin{align}\label{eq:f_res_decorrelated}
    f_{n + 1}^{\textrm{in}} &= f_1^0(\vector{q}_k, -\vector{p}_k) f_n \textrm{,}
\end{align}
where $f_1^0$ is the equilibrium single-particle density of the reservoir, which is hence assumed to be statistically independent of the open system. Another model established in ref. \cite{delle_site_liouville-type_2020} would be 
\begin{align}\label{eq:f_res_grandcanonical}
    f_{n + 1}^{\textrm{in}} &= f_{n + 1}^{\textrm{GC}} \textrm{,}
\end{align}
imposing a grand-canonical distribution $f_{n + 1}^{\textrm{GC}}$ for the reservoir.

\subsection{Correspondence between the Liouville-type hierarchy and the AdResS method}
It is emphasized that the Liouville-type hierarchy is concerned with the open system and its boundary, which in the AdResS method would correspond to the \textsc{AT} region and the \textsc{AT/$\Delta$} boundaries, respectively. This means that, within the hierarchy, there is no conception of either the \textsc{$\Delta$/TR} interfaces or the \textsc{TR} region altogether. The \textsc{$\Delta$} regions can thus be regarded as a representation of the reservoir responsible for creating the correct statistics of particles at the open system's boundary in the sense of the coupling terms of Eqs. (\ref{eq:psi}) and (\ref{eq:phi}). The \textsc{TR} region then only serves the purpose of providing the particles necessary to produce that statistics and the \textsc{$\Delta$/TR} interface corrections are the means by which the particles equilibrate within the \textsc{$\Delta$} regions. 

These considerations will inspire a novel interpretation of the thermodynamic force in terms of the Liouville-type hierarchy for the open system as discussed in the next section. 
\section{The thermodynamic force interpreted in terms of the Liouville-type hierarchy}
From the correspondence between the Liouville-type hierarchy and the AdResS method we can conclude that in order for the \textsc{AT} region to be a physically consistent representation of the open system, the AdResS reservoir needs to provide conditions at the \textsc{AT/$\Delta$} boundaries that are in accord with the source terms $\Psi_n$ and $\Phi_n^{n + 1}$. It is apparent that the boundaries of an open subsystem in a full-atomistic reference simulation naturally adhere to this condition. 

In particular, this means that the average force $\vector{F}_{\textrm{av}}$ established in Eq. (\ref{eq:avForce}) that is acting through the \textsc{AT/$\Delta$} boundary at $x_{\textrm{AT/$\Delta$}}$ on a particle $i$ at position $\vector{q}_i$ residing in the \textsc{AT} region needs to be equal to the average force through a boundary on a particle at the same respective positions in the full-atomistic reference simulation, meaning
\begin{align}\label{eq:F_av_AdResS_vs_FullAT}
    \vector{F}_{\textrm{av, AdResS}}(\vector{q}_i; x_{\textrm{AT/$\Delta$}}) &= \vector{F}_{\textrm{av, Full-AT}}(\vector{q}_i; x_{\textrm{AT/$\Delta$}}) = \vector{F}_{\textrm{av}}(\vector{q}_i; x_{\textrm{AT/$\Delta$}})
\end{align}
Due to the left- and right-hand boundaries being statistically equivalent, we will from now on treat only the right-hand \textsc{AT/$\Delta$} boundary and \textsc{$\Delta$/TR} interface at $x_{\textrm{AT/$\Delta$}}$ and $x_{\textrm{$\Delta$/TR}}$, respectively, without loss of generality. 

\begin{figure}[h]
    \centering
    \input{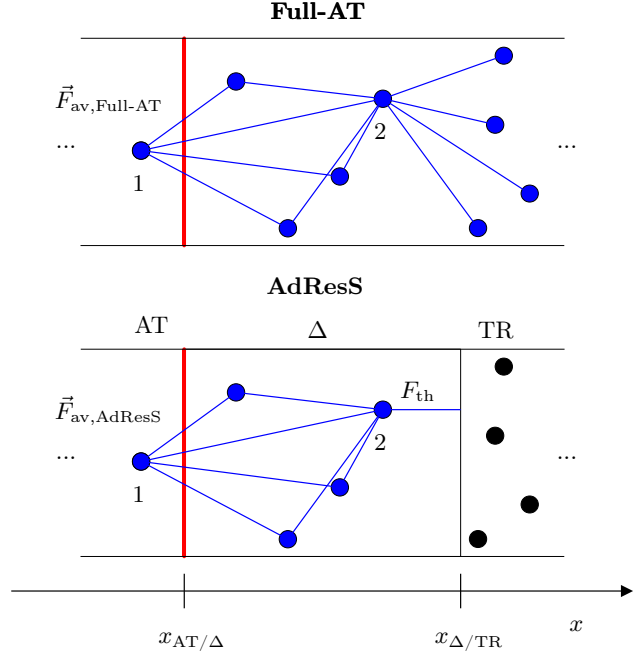}
    \vspace{-1.2cm} 
    \caption{Situation at the open system's boundary with particle $1$ residing within the open system and particle $2$ residing in the reservoir. Full-\textsc{AT}: Particle $1$ experiences an average force $\vector{F}_{\textrm{av, Full-AT}}$ through the boundary at $x_{\textrm{AT/$\Delta$}}$ due to the particles in the reservoir, which are correctly distributed due to themselves having a correct environment (see particle $2$). AdResS: Particle $1$ experiences an average force $\vector{F}_{\textrm{av, AdResS}}$ through the \textsc{AT/$\Delta$} boundary. However, the particles in \textsc{$\Delta$}, e.g. particle $2$, lack the forces through the \textsc{$\Delta$/TR} interface at $x_{\textrm{$\Delta$/TR}}$. These, among other effects, must be reintroduced in an average fashion by the thermodynamic force $F_{\textrm{th}}$.}
    \label{fig:illustration_F_av_adress_vs_FAT}
\end{figure}

In order for Eq. (\ref{eq:F_av_AdResS_vs_FullAT}) to hold, the integral in Eq. (\ref{eq:avForce}) evaluated from an AdResS simulation must agree with the same integral evaluated from the full-atomistic reference. This is obviously the case if the two-point-statistics $f_{k|i}^0$ measured through the \textsc{AT/$\Delta$} boundary in the two simulations also agree with each other. However, the particles residing in the \textsc{$\Delta$} region of AdResS themselves lack any force with the particles in the \textsc{TR} region (see Fig. \ref{fig:illustration_F_av_adress_vs_FAT}). In this strongly artificial environment the \textsc{$\Delta$} particles can, in general, not be expected to behave close to fully atomistic and produce the required $f_{k|i}^0$ for the particles in \textsc{AT}.  

This artifact of the \textsc{$\Delta$/TR} interface has to be remedied on a mean-field level by the thermodynamic force, effectively reintroducing the average force $\vector{F}_{\textrm{av}}(\vector{q}_i; x_{\textrm{$\Delta$/TR}})$ that particles in \textsc{$\Delta$} would feel through a boundary at $x_{\textrm{$\Delta$/TR}}$ if the interactions had not been switched off. It is thus clear that $\vector{F}_{\textrm{av}}(\vector{q}_i; x_{\textrm{$\Delta$/TR}})$ comprises one contribution to the thermodynamic force and a close resemblance between these two forces may be expected within the \textsc{$\Delta$} region. Since the other interface artifacts have not been considered explicitly in this argument, deviations between the two forces close to the \textsc{$\Delta$/TR} interface and in the \textsc{TR} region are also to be expected. 

The extent of this resemblance and the implications for both the Liouville-type hierarchy and the AdResS method can be better understood with a more explicit expression for $\vector{F}_{\textrm{av}}$ at hand, which we will hence derive in this section. 
In that, we will first derive a general expression for $\vector{F}_{\textrm{av}}$ at the open system's boundary and then apply it to the situation at the \textsc{$\Delta$/TR} interface. 

Our derivation is similar in spirit to the works of Werder \textit{et al.}\cite{werder_hybrid_2005} and Kotsalis \textit{et al.}\cite{kotsalis_control_2007}, who obtained a double-integral expression for an effective boundary force in a molecular dynamics simulation of an incompressible Lennard-Jones fluid (i.e. with constant number of particles). We emphasize that here we derive a single-integral expression for $F_{\textrm{av}}$, which facilitates the interpretation of the thermodynamic force from the perspective of the Liouville-type hierarchy and that we are concerned explicitly with systems open for the exchange of particles with a reservoir.

\subsection{Simplifying the integrand}
Assuming \emph{global thermal equilibrium} for the universe $U$ and using Bayes' rule, the conditional probability density in Eq. \ref{eq:avForce} can be rewritten as (see Eq. (3.3) on page 10 in ref. \cite{van_kampen_stochastic_2007})
\begin{align}\label{eq:joint_distribution}
    f_{k|i}^0(\vector{q}_k, \vector{p}_k|\vector{q}_i, \vector{p}_i) &= \frac{f_2^0(\vector{q}_i, \vector{q}_k, \vector{p}_i, \vector{p}_k)}{f_1^0(\vector{q}_i, \vector{p}_i)} \textrm{,}
\end{align}
where $f_2^0$ is the equilibrium two-particle joint probability density of finding particles $i$ and $k$ in states $(\vector{q}_i, \vector{p}_i)$ and $(\vector{q}_k, \vector{p}_k)$, respectively, and $f_1^0$ is the equilibrium single-particle probability density of the system and equal to the assumed reservoir density introduced in Eq. (\ref{eq:f_res_decorrelated}).

For a \emph{simple atomistic fluid}, these one- and two-particle probability densities can be factorized into configurational and momentum parts and the momentum parts further factorize into independent one-particle Maxwell-Boltzmann distributions $f_{\textrm{M}}(p_i)$ depending only on the magnitude of the respective momentum $p_i = \abs{\vector{p}_i}$ as (see Eqs. (4.2.2) and (4.2.3) on page 110 of ref. \cite{hansen_theory_2013})
\begin{subequations}\label{eq:probability_density_factorization}
\begin{align}
    f_2^0(\vector{q}_i, \vector{q}_k, \vector{p}_i, \vector{p}_k) &= \rho^{(2)}(\vector{q}_i, \vector{q}_k) f_{\textrm{M}}(p_i) f_{\textrm{M}}(p_k)\\
    f_1^0(\vector{q}_i, \vector{p}_i) &= \rho^{(1)}(\vector{q}_i) f_{\textrm{M}}(p_i) \textrm{.}
\end{align}
\end{subequations}

The configurational parts are the well-known one- and two-particle densities, defined for a \emph{homogeneous} and \emph{isotropic} fluid as (see eqs. (2.5.6) and (2.5.9) on p. 33 in ref. \cite{hansen_theory_2013})
\begin{subequations}\label{eq:particle_density_factorization}
\begin{align}
    \rho^{(2)}(\vector{q}_i, \vector{q}_k) &= \rho^2 g(r) \\
    \rho^{(1)}(\vector{q}_i) &= \rho \textrm{,}
\end{align}
\end{subequations}
where $\rho$ and $g(r)$ are the particle number density and the radial distribution function of the equilibrated universe, respectively, and $r = |\vector{q}_i - \vector{q}_k|$ is the distance between particles $i$ and $k$. 

Eq. (\ref{eq:joint_distribution}) thus reduces to   
\begin{align}
    f_{k|i}^0(\vector{q}_k, \vector{p}_k|\vector{q}_i, \vector{p}_i) &= f_{k|i}^0(\vector{q}_k, \vector{p}_k|\vector{q}_i) = \rho g(r) f_{\textrm{M}}(\vector{p}_k) \textrm{.}
\end{align}
For a radially dependent pair potential $V(\vector{q}_i - \vector{q}_k) = V(r)$, the average force integral may be rewritten as 
\begin{align}
    \vector{F}_{\textrm{av}}(\vector{q}_i; \partial \Omega) &= -\rho \int_{\Omega_{\textrm{c}}} \dd{\vector{q}_k} \nabla_{\vector{q}_i} V(r) g(r) \int_{\R^3} \dd{\vector{p}_k} f_{\textrm{M}}(\vector{p}_k) = -\rho \int_{\Omega_{\textrm{c}}} \dd{\vector{q}_k} \nabla_{\vector{q}_i} V(r) g(r) \textrm{.}
\end{align}
It will turn out to be convenient to transform the gradient into spherical polar coordinates and to consider the force components as projections of the radial force onto the respective cartesian axis. Hence, we write
\begin{align}\label{eq:general_F_av}
    \vector{F}_{\textrm{av}}(\vector{q}_i; \partial \Omega) &= -\rho \int_{\Omega_{\textrm{c}}} \dd{\vector{q}_k} \frac{(\vector{q}_i - \vector{q}_k)}{r} \partofwrt{V(r)}{r} g(r) \textrm{.}
\end{align}

\subsection{Simplifying the integration domain}
For the purpose of this study, we now assume the boundary $\partial \Omega$ to have \emph{planar geometry} and place it parallel to the $yz$-plane at a position $x_{\partial}$. The integration domain of Eq. (\ref{eq:general_F_av}) is then given as 
\begin{align}
    \Omega_{\textrm{c}} &= [x_{\partial}, \infty) \cross \R^2 \textrm{.}
\end{align}
Since we considered the system and reservoir to be homogeneous and isotropic, there can only be non-zero contributions to $\vector{F}_{\textrm{av}}$ in the surface-normal direction $x$, and it can only depend on the left-hand distance $r_{\partial} = x_{\partial} - x_i \geq 0$ to the boundary, where $x_i$ is the position of the particle $i$ in the $x$-direction. This enables us to simplify the force according to 
\begin{align}\label{eq:F_av_in_terms_of_F_av_x}
    \vector{F}_{\textrm{av}}(\vector{q}_i; x_{\partial}) &= \vector{F}_{\textrm{av}}(r_{\partial}) = \posVector{1}{0}{0} F_{\textrm{av}}^x(r_{\partial}) \textrm{.}
\end{align}

Combining with Eq. (\ref{eq:general_F_av}), we obtain
\begin{align}
    F_{\textrm{av}}^x(r_{\partial}) &= -\rho \int_{\Omega_{\textrm{c}}} \dd{\vector{q}_k} \frac{r_x}{r} \partofwrt{V(r)}{r} g(r) \textrm{.}
\end{align}
where $r_x = x_i - x_k$.

\begin{figure}[ht]
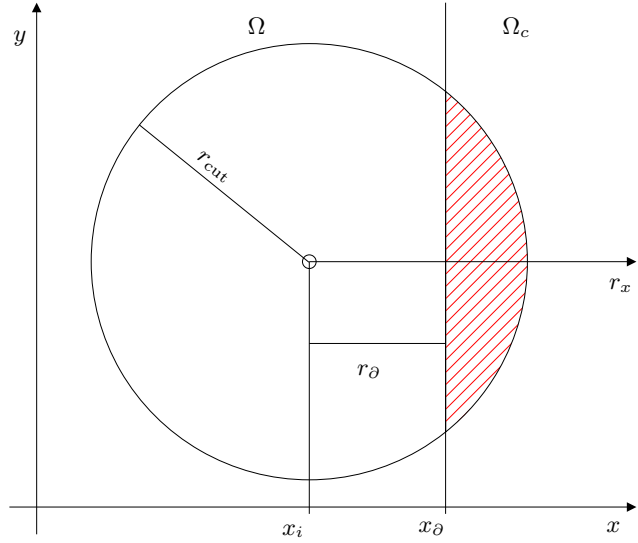

    \centering
    \include{Figures/integration_domain_tex.tex}
    \caption{Illustration of the integral appearing in Eq. (\ref{eq:F_av_x}), projected into the $xy$ plane. The simplified integration domain is shaded red and was obtained by transforming into spherical polar coordinates with polar axis $r_{x}$, a potential with cut-off radius $r_{\textrm{cut}}$ and the dependence on the distance $r_{\partial}$ of the particle $i$ at position $x_i$ to the boundary at position $x_{\partial}$. The domain of the open system where the particle resides is $\Omega$ and the reservoir region is $\Omega_c$.}
    \label{fig:V_av_domain}
\end{figure}

The integration can be carried out conveniently in spherical polar coordinates with radius $r$, polar angle $\varphi$ and azimuthal angle $\theta$, volume element $\dd{\vector{q}_k} = \dd{r}\dd{\varphi}\dd{\theta} r^2 \sin{\varphi}$ and the polar axis $r_x = r \cos{\varphi}$ aligned with the $x$-axis of the Cartesian coordinate system (see Fig. \ref{fig:V_av_domain}). Then, 
\begin{align}
    F_{\textrm{av}}^x(r_{\partial}) &= -\rho \int_{r_{\textrm{min}}}^{r_{\textrm{max}}} \dd{r} \int_{\varphi_{\textrm{min}}}^{\varphi_{\textrm{max}}} \dd{\varphi} \int_{0}^{2\pi} \dd{\theta} r^2 \sin{\varphi}\cos{\varphi} \partofwrt{V(r)}{r} g(r) \nonumber\\
    &= -2\pi \rho \int_{r_{\textrm{min}}}^{r_{\textrm{max}}} \dd{r} r^2 \partofwrt{V(r)}{r} g(r) \int_{\varphi_{\textrm{min}}}^{\varphi_{\textrm{max}}} \dd{\varphi} \sin{\varphi}\cos{\varphi} \textrm{,}
\end{align}
where $\theta$ was integrated out due to the rotational symmetry of the problem around the polar axis $r_x$. The lower limit of integration over the polar angle is $\varphi_{\textrm{min}} = 0$ while the upper limit depends on the radius as $\varphi_{\textrm{max}} = \arccos{\frac{r_{\partial}}{r}}$. Carrying out the integration over $\varphi$ then yields
\begin{align}
    F_{\textrm{av}}^x(r_{\partial}) &= -\pi \rho \int_{r_{\textrm{min}}}^{r_{\textrm{max}}} (r^2 - r_{\partial}^2) \partofwrt{V(r)}{r} g(r) \dd{r} \textrm{.}
\end{align}

Obviously, the lower limit for the radial integration needs to be the distance between the particle and the interface $r_{\textrm{min}} = r_{\partial}$, whilst the upper limit is the cut-off distance $r_{\textrm{max}} = r_{\textrm{cut}}$, resulting in
\begin{align}
    F_{\textrm{av}}^x(r_{\partial}) &= -\pi \rho \int_{r_{\partial}}^{r_{\textrm{cut}}} (r^2 - r_{\partial}^2) \partofwrt{V(r)}{r} g(r) \dd{r}.
\end{align}
This can be reorganized in terms of the integral expression 
\begin{align}\label{eq:I_n_integral_expression}
    I_n(r_{\partial}) &= \int_{r_{\partial}}^{r_{\textrm{cut}}} \dd{r} r^n \partofwrt{V}{r} g(r) 
\end{align}
as 
\begin{align}\label{eq:F_av_x}
    F_{\textrm{av}}^x(r_{\partial}) &= \pi \rho (r_{\partial}^2 I_0 - I_2) \textrm{.}
\end{align}

Reinserting Eq. (\ref{eq:F_av_x}) into Eq. (\ref{eq:F_av_in_terms_of_F_av_x}), one arrives at the following expression for the average force acting on particle $i$ close to the boundary 
\begin{align}\label{eq:F_av_explicit}
    \vector{F}_{\textrm{av}}(r_{\partial}) &= \pi \rho \posVector{1}{0}{0} (r_{\partial}^2 I_0 - I_2) \textrm{.}
\end{align}

This force is conservative and the corresponding potential can be readily calculated as 
\begin{align}\label{eq:V_av_explicit}
    V_{\textrm{av}}(r_{\partial}) &= \frac{\pi \rho}{3} (r_{\partial}^3 I_0 - 3 r_{\partial} I_2 + 2 I_3) \textrm{.}
\end{align}

\subsection{Significance of the average force for the AdResS thermodynamic force}
We are now in a position to understand the significance of the average force for the thermodynamic force. We carry on with our discussion at the beginning of this section by noting that the lacking average force of the \textsc{$\Delta$} region is $F_{\textrm{av}}^x(r_{\textrm{$\Delta$/TR}})$ of Eq. (\ref{eq:F_av_x}), where $r_{\textrm{$\Delta$/TR}} = x_{\textrm{$\Delta$/TR}} - x \geq 0$, effectively placing the ``boundary'' at the position of the \textsc{$\Delta$/TR} interface, $x_{\textrm{$\Delta$/TR}}$. We dropped the particle index $i$ for the $x$-coordinate to simplify the comparison with the thermodynamic force. 

From the correspondence of the Liouville-type hierarchy and the AdResS method, we concluded that the statistics and average force through the \textsc{AT/$\Delta$} boundary at position $x_{\textrm{AT/$\Delta$}}$ need to be approximately equal to their full-atomistic equivalents, which yielded Eq. (\ref{eq:F_av_AdResS_vs_FullAT}). Incorporating the geometry of our domain, this means that
\begin{align}\label{eq:F_av_x_AdResS_vs_FAT}
    F_{\textrm{av, AdResS}}^x(x_{\textrm{AT/$\Delta$}} - x) &= F_{\textrm{av, Full-AT}}^x(x_{\textrm{AT/$\Delta$}} - x) = F_{\textrm{av}}^x(x_{\textrm{AT/$\Delta$}} - x) \textrm{.}
\end{align}

We already pointed out that in order for the \textsc{$\Delta$} particles to be sufficiently equilibrated at the \textsc{AT/$\Delta$} boundary, the thermodynamic force must reintroduce the lacking average force $F_{\textrm{av}}^x$, leading us to conclude that
\begin{align}\label{eq:F_th_decomposition_into_F_av_and_F_dash}
    F_{\textrm{th}}(x) &= F_{\textrm{av}}^x( x_{\textrm{$\Delta$/TR}} - x) + \hat{F}(x) \textrm{,}
\end{align}
where $\hat{F}$ is a, yet unknown, deviation of the thermodynamic force from the lacking average force due to the other artifacts of the \textsc{$\Delta$/TR} interface introduced in section \ref{sec:interface_corrections}. Among these are the residual effects of the excluded volume overlap at the \textsc{$\Delta$/TR} interface, which are in parts compensated by the thermostat and the force-capping scheme. We thus expect $\hat{F}$ to have a parametric dependence on the thermostat coupling parameter $\gamma$ and the capping radius $r_{\textrm{cap}}$ ($\hat{F} = \hat{F}(x; \gamma, r_{\textrm{cap}})$).

If the combined thermostat coupling and force-capping scheme are sufficiently strong and the thermodynamic differences between \textsc{AT} and \textsc{TR} can be compensated close to the \textsc{$\Delta$/TR} interface such that $\hat{F}$ goes to zero at a position $x_{\textrm{$\Delta$/TR}} \geq x_{\textrm{fade}} \geq x_{\textrm{AT/$\Delta$}}$ within the \textsc{$\Delta$} region, the thermodynamic force in that region can be written as
\begin{align}\label{eq:F_th_close_to_boundary}
    F_{\textrm{th}}(x) &= F_{\textrm{av}}^x(x_{\textrm{$\Delta$/TR}} - x) \textrm{, for } x_{\textrm{AT/$\Delta$}} \leq x \leq x_{\textrm{fade}} \leq x_{\textrm{$\Delta$/TR}} \textrm{.}    
\end{align} 
It should be emphasized that this is merely an assumption. However, this is to be understood in the following sense: If there is no $x_{\textrm{fade}}$ meaning that $\hat{F}$ does not vanish within \textsc{$\Delta$} and Eq. (\ref{eq:F_th_close_to_boundary}) does not hold true, the \textsc{$\Delta$/TR} interface artifacts propagate into the \textsc{AT} region, where they can not be compensated by the thermodynamic force. This would result in the \textsc{AT} region not being a physically consistent representation of an open system. Conversely, if we observe during the preliminary phase of an AdResS simulation that the converged thermodynamic force obtained from the iterative procedure of Eq. (\ref{eq:thermo_force_iterations}) matches with $F_{\textrm{av}}^x$ within $x_{\textrm{AT/$\Delta$}} \leq x \leq x_{\textrm{fade}}$, we know that the only artifact left to compensate close to the \textsc{AT/$\Delta$} boundary is the missing average force. $F_{\textrm{av}}^x(r_{\textrm{$\Delta$/TR}})$ smoothly fades to zero for $r_{\textrm{$\Delta$/TR}}$ approaching a distance of one cut-off radius $r_{\textrm{cut}}$ to the \textsc{$\Delta$/TR} interface. It is customary to use a width $L_{\Delta} = r_{\textrm{cut}}$ of the \textsc{$\Delta$} region in AdResS simulations\cite{delle_site_molecular_2019}. The thermodynamic force then also smoothly goes to zero at $x_{\textrm{AT/$\Delta$}}$ and the \textsc{AT} region is physically consistent. 

Assuming now that Eq. (\ref{eq:F_th_close_to_boundary}) indeed holds true, we can derive the properties of the thermodynamic force that we expect in the \textsc{$\Delta$} region close to the \textsc{AT/$\Delta$} boundary, where $x_{\textrm{AT/$\Delta$}} \leq x \leq x_{\textrm{fade}}$. 

\subsection{Properties of the thermodynamic force close to the boundary}\label{subsec:properties_of_thermodynamic_force}
From Eq. (\ref{eq:F_av_x}), it is clear that $F_{\textrm{av}}^x$ has a parametric dependence on the cut-off radius $r_{\textrm{cut}}$ and the density $\rho$ and a functional dependence on the interparticle potential $V(r)$ and the radial distribution function $g(r)$, which itself is known to be a unique function of the density $\rho$, temperature $T$ and the potential (see section 4.7.4 on p. 174 of ref. \cite{tuckerman_statistical_2023}).

These points can be taken as strong hints that at least the part of the thermodynamic force close to the \textsc{AT/$\Delta$} boundary is completely determined by 
\begin{enumerate}
    \item the thermodynamic state represented by $\rho$ and $T$,
    \item the fluid model given by $V$,
    \item the distance $r_{\textrm{$\Delta$/TR}}$ to the \textsc{$\Delta$/TR} interface.
\end{enumerate}

Perhaps even more interesting are the properties that the thermodynamic force close to the \textsc{AT/$\Delta$} boundary does not depend on under the assumption of Eq. (\ref{eq:F_th_close_to_boundary}). 

Firstly, the expression in Eq. (\ref{eq:F_av_x}) was derived making use of infinite planar boundaries for the open system. This is well-reflected in the fact that, for the slab-shaped AdResS simulation setup, the box is periodic in the $y$ and $z$ directions. Consequently, the thermodynamic force close to the \textsc{AT/$\Delta$} boundary is expected to be independent of the edge lengths $L_y$ and $L_z$ of the periodic box in the $y$ and $z$ directions, given that it is large enough to produce the correct statistics and not suffer from prevalent finite-size effects.

Furthermore, $F_{\textrm{av}}^x$ and thus $F_{\textrm{th}}$ close to the \textsc{AT/$\Delta$} boundary is completely independent of any properties of the \textsc{TR} region, such as the width of the region $L_{\textrm{TR}}$ or the chosen model for the \textsc{TR} particles, as long as the \textsc{TR} region can provide a sufficient supply of particles.

Along the same lines we can say that $F_{\textrm{th}}$ does not depend on the width $L_{\Delta}$ of the \textsc{$\Delta$} region as long as the \textsc{AT/$\Delta$} boundary is at least one cut-off length away from the \textsc{$\Delta$/TR} interface. This is due to $F_{\textrm{av}}^x$ smoothly going to zero at that distance and then remaining zero beyond it. 

$F_{\textrm{av}}^x$ is also independent of the width $L_{\textrm{AT}}$ of the \textsc{AT} region, provided that the \textsc{$\Delta$} particles are subject to a full-atomistic environment at and through the \textsc{AT/$\Delta$} boundary. 

Another important aspect to be noted is that $F_{\textrm{av}}^x$ as seen in Eq. (\ref{eq:F_av_x}) does not depend on momenta. Thus, the thermodynamic force close to the \textsc{AT/$\Delta$} boundary is expected to be independent of the coupling parameter $\gamma$ of the applied thermostat. 

$F_{\textrm{av}}^x$ also only implicitly depends on the capping radius $r_{\textrm{cap}}$ introduced in Eq. (\ref{eq:V_cut_cap}) through the potential $V$. However, as the capping scheme changes the potential in its strongly repulsive regime at small distances and the radial distribution function is zero there, this effect has no consequences for the value of the integral in Eq. (\ref{eq:F_av_x}), rendering the thermodynamic force close to the \textsc{AT/$\Delta$} boundary independent of $r_{\textrm{cap}}$. 

We can then summarize that under the conditions pointed out in the preceding paragraphs and if $\hat{F}$ fades sufficiently fast so that Eq. (\ref{eq:F_th_close_to_boundary}) holds true, the thermodynamic force close to the \textsc{AT/$\Delta$} boundary is independent of 
\begin{enumerate}\label{enum:F_th_independencies}
    \item the box dimensions $L_y$ and $L_z$ in $y$ and $z$ directions,
    \item the properties of the \textsc{TR} region, such as its width $L_{\textrm{TR}}$,
    \item the width $L_{\Delta}$ of the \textsc{$\Delta$} region,
    \item the width $L_{\textrm{AT}}$ of the \textsc{AT} region,
    \item the thermostat coupling parameter $\gamma$,
    \item the capping radius $r_{\textrm{cap}}$.
\end{enumerate}

\subsection{Implications for the calculation of the thermodynamic force}\label{subsec:implications_for_the_thermodynamic_force}
From the conclusions we drew in section \ref{subsec:properties_of_thermodynamic_force}, we can derive the following simplifications for the calculation of the thermodynamic force:
\begin{itemize}
    \item Initial guess - if we know the thermodynamic state variables $\rho$ and $T$, the fluid model $V$ and a closed form of the radial distribution function $g(r)$, we can in principle evaluate the integral in Eq. (\ref{eq:F_av_x}). Together with Eq. (\ref{eq:F_th_close_to_boundary}), this comprises an \emph{a priori} approximation to the thermodynamic force that may enter into the iterative procedure of Eq. (\ref{eq:thermo_force_iterations}) as an educated initial guess. As this is anyways an approximation, an approximate form of $g(r)$ would already be sufficient for that purpose. 
    \item No \textsc{AT} region - a \textsc{$\Delta$} region provides through the respective \textsc{AT/$\Delta$} boundary a full-atomistic environment for the particles within the nearby \textsc{AT} region. The particles within the \textsc{$\Delta$} region themselves need to experience a full-atomistic environment through the \textsc{AT/$\Delta$} boundary in order to properly equilibrate. Consequently, it stands to reason that one could create this environment with just another \textsc{$\Delta$} region on the other side of the boundary. This would effectively mean getting rid of the entire \textsc{AT} region altogether and placing the left- and right-hand \textsc{AT/$\Delta$} boundaries on top of each other, substantially decreasing the computational cost for determining the thermodynamic force. 
    \item Minimal size in $y$ and $z$ - the size $L_y$ and $L_z$ of the periodic unit box in $y$ and $z$ directions can be reduced to a minimum, the size of which remains to be determined.  
\end{itemize}
In addition to the simplifications proposed above, one can also employ quality criteria for the consistency of the converged thermodynamic force that are directly linked to the reservoir coupling terms appearing in the Liouville-type hierarchy in Eqs. (\ref{eq:psi}) and (\ref{eq:phi}):
\begin{itemize}
    \item Particle number density at the \textsc{AT/$\Delta$} boundary - from the assumption made in Eq. (\ref{eq:f_res_decorrelated}) we can conclude that the reservoir needs to provide the correct equilibrium probability density $f_1^0$. Following our treatment of this function for the combined system and reservoir in Eqs. (\ref{eq:probability_density_factorization}) and (\ref{eq:particle_density_factorization}) we know that this requires the particles entering into the system to have the correct one-particle density $\rho(x_{\textrm{AT/$\Delta$}}) = \rho$.   
    \item Maxwell-Boltzmann momentum distribution at the \textsc{AT/$\Delta$} boundary - on the same note as the last point, we see from Eq. (\ref{eq:particle_density_factorization}) that the particles entering into the system need to have the Maxwell-Boltzmann momentum distribution $f_{\textrm{M}}(p; x_{\textrm{AT/$\Delta$}}) = f_{\textrm{M}}(p)$.
    \item Statistics of particle crossings through the \textsc{AT/$\Delta$} boundary - if we consider the modeling assumption of Eq. (\ref{eq:f_res_grandcanonical}) for the influx of probability density through the boundary, we can conclude that the reservoir would not only need to provide the correct equilibrium average density, but also fluctuations thereof. A physically sound environment for an AdResS simulation would thus need to exhibit the correct fluctuations in the net particle flux through the \textsc{AT/$\Delta$} boundary, which can be measured via the net number of particle crossings. We elaborate this point further in appendix \ref{app:net_flux}.   
    \item Average force through the \textsc{AT/$\Delta$} boundary - from Eq. (\ref{eq:F_av_x_AdResS_vs_FAT}) it is clear that the average force through the interface $F_{\textrm{av}}^x(x_{\textrm{AT/$\Delta$}} - x)$ measured in an AdResS simulation must match the full-atomistic reference. We elaborate this point further in appendix \ref{app:F_av_x}. 
\end{itemize}

We can see that all quality criteria are properties of the \textsc{AT/$\Delta$} boundary and the \textsc{AT} region does not appear explicitly. This further illustrates that the thermodynamic force is indeed a consequence of the artificial \textsc{$\Delta$/TR} interface. 

In summary, we suggest here a minimal version of the AdResS method that can be employed for the iterative calculation of the thermodynamic force. We showed that this scheme can be justified from considerations of and validated with observables inspired by the Liouville-type hierarchy for an open system. This serves both as a corroboration of the physical soundness of the AdResS method and as a cross-check for the assumptions made in the construction of the Liouville-type hierarchy.

\begin{figure}[ht]
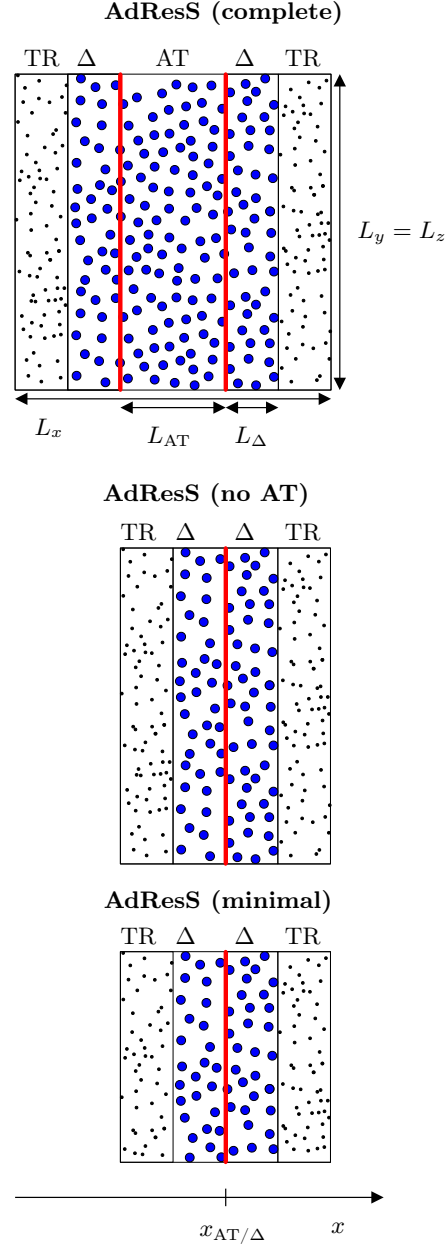

    \centering
    \hspace{2.5cm}\include{Figures/complete_noAT_minimal_tex.tex}
    \vspace{-1.3cm}
    \caption{Proposed box setups for the calculation and validation of the thermodynamic force. AdResS (complete): AdResS setup as it has been used so far with an extended \textsc{AT} region, two adjacent \textsc{$\Delta$} regions and a \textsc{TR} region beyond two \textsc{$\Delta$/TR} interfaces. AdResS (no \textsc{AT}): Like complete but with $L_{\textrm{AT}} = 0$, effectively collapsing the two \textsc{AT/$\Delta$} boundaries onto a single one. AdResS (minimal): Like no \textsc{AT}, but with smaller box dimensions $L_y$ and $L_z$.}
    \label{fig:AdResS_vs_minimalAdResS}
\end{figure}

The proposed minimal AdResS scheme without \textsc{AT} region and with minimal box dimensions $L_y$ and $L_z$ is depicted in Fig. \ref{fig:AdResS_vs_minimalAdResS} alongside an intermediate scheme where only the \textsc{AT} region is omitted (no \textsc{AT}) and the regular AdResS setup (complete). If Eq. (\ref{eq:F_th_close_to_boundary}) holds true, any of these schemes is expected to yield the same thermodynamic force and obtain the same statistics at the respective \textsc{AT/$\Delta$} boundary. In the case of the minimal and no \textsc{AT} setups, we will henceforth denote the boundary in between the two adjacent \textsc{$\Delta$} regions as the \textsc{AT/$\Delta$} boundary as well. 

Of course, a complete AdResS simulation including an \textsc{AT} region will still need to be run at production level in order to obtain properties of the open system. However, this can easily be set up by calculating the thermodynamic force with minimal AdResS and then running a production simulation where one applies the calculated thermodynamic force at the respective \textsc{$\Delta$/TR} interfaces of a complete AdResS box. 

The claims made in this section are qualitative statements based on theoretical considerations. Quantitative support by means of a numerical simulation study will thus be documented in the next section. 
\section{Numerical Experiments}\label{sec:numerical_experiments}
Molecular dynamics simulations were carried out to test the validity of the claims made in sections \ref{subsec:properties_of_thermodynamic_force} and \ref{subsec:implications_for_the_thermodynamic_force}. For these, we used the AdResS scheme with abrupt \textsc{$\Delta$/TR} interface described in section \ref{sec:simulation_method_and_theory} and the truncated and shifted Lennard-Jones potential for the \textsc{AT} and \textsc{$\Delta$} regions. This amounts to inserting  
\begin{align}
    v(r) &= v_{\textrm{LJ}}(r) - v_{\textrm{LJ}}(r_{\textrm{cut}})
\end{align}
with 
\begin{align}
    v_{\textrm{LJ}}(r) &= 4\epsilon \left[\left(\frac{\sigma}{r}\right)^{12} - \left(\frac{\sigma}{r}\right)^6\right]
\end{align}
into Eq. (\ref{eq:V_cut_cap}). Here, $\epsilon$ and $\sigma$ are the minimum energy and the root distance of the Lennard-Jones potential, respectively. We carried out our simulations in reduced Lennard-Jones units ($\sigma = \epsilon = m = 1)$ and with a cut-off radius of $r_{\textrm{cut}} = \SI{2.5}{\sigma}$ and a capping radius of $r_{\textrm{cap}} \approx \SI{0.82}{\sigma}$.

We implemented the required functionalities in the open-source software package Multi-Resolution Molecular Dynamics (\textsc{MRMD}) available at GitHub\cite{eibl_multi_2025}. The details of those simulations are reported in appendix \ref{app:simulation_details}. We chose to investigate the influence of the following parameters: The temperature $T$, the density $\rho$, the width $L_{\textrm{AT}}$ of the \textsc{AT} region, the box lengths $L_{y, z}$ in $y$ and $z$ directions. Furthermore, we tested the suitability of using $F_{\textrm{av}}^x$ as an initial guess. The protocol which we followed in our simulations is reported in the next subsection. 

\subsection{Simulation protocol}\label{subsec:simulation_protocol}
In order to test for the influence of temperature and density, we conducted simulations for three different thermodynamic states: (1) low density and low temperature, (2) high density and high temperature and (3) high density and high temperature, where the low density was $\rho = \SI{0.296}{\sigma^{-3}}$, the high density was $\rho = \SI{0.370}{\sigma^{-3}}$, the low temperature was $T = \SI{1.5}{\epsilon/k_{\textrm{B}}}$ and the high temperature was $T = \SI{2.0}{\epsilon/k_{\textrm{B}}}$. These thermodynamic states are all well above the critical curve for the truncated and shifted Lennard-Jones potential\cite{smit_phase_1992}.

For each of those conditions, we performed simulations using the three box setups introduced in fig. \ref{fig:AdResS_vs_minimalAdResS}: (1) AdResS (minimal), (2) AdResS (no \textsc{AT}), (3) AdResS (complete) (see app. \ref{app:simulation_details} for details). All our simulations followed the three-step protocol given in tab. \ref{tab:simulation_protocol}.
\begin{table}[ht]
    \caption{\label{tab:simulation_protocol}Simulation protocol.}
    \setlength{\arrayrulewidth}{0.2mm}
    \setlength{\tabcolsep}{6pt}
    \renewcommand{\arraystretch}{1.5}
    \centering
    \begin{tabular}{ l l }
        \hline
        \textbf{1} & Calculation of the thermodynamic force initial guess \\
        \textbf{2} & Iterative calculation of the thermodynamic force \\
        \textbf{3} & Validation of the calculated thermodynamic force \\
        \hline
    \end{tabular}
\end{table}

During the validation step, we additionally performed a full-atomistic reference simulation for each thermodynamic state using the box dimensions $L_{x, y, z} = \SI{30.0}{\sigma}$ and the periodic boundary conditions of the complete setup. 

The results of these simulations are elaborated in the next subsections.

\subsection{Calculation of the thermodynamic force initial guess}\label{subsec:initial_guess}
As has been stated in section \ref{subsec:implications_for_the_thermodynamic_force}, the thermodynamic force close to the \textsc{AT/$\Delta$} boundary can be calculated \emph{a priori} if the thermodynamic variables $\rho$ and $T$, the fluid model potential $V(r)$ and the radial distribution function $g(r)$ are known. Temperature, density and potential are chosen in the equilibration stage of the molecular dynamics simulation and are therefore known. 

Conveniently, during the preparation of an AdResS simulation, one has to equilibrate the system under study. One can thus measure $g(r)$ before carrying out the iterative thermodynamic force calculation, albeit with noise, and use it to obtain an approximation to $F_{\textrm{av}}^x$ of Eq. (\ref{eq:F_av_x}). This approximation can then enter in the iterative procedure as an educated initial guess according to the points made in section \ref{subsec:implications_for_the_thermodynamic_force}. In the special case of the Lennard-Jones fluid, one can alternatively obtain such an initial guess by using a parametrized form of $g(r)$ such as the ones provided by Morsali \textit{et al.} or Goldman\cite{morsali_accurate_2005, goldman_explicit_1979}. 

\begin{figure}[ht]
    \centering
    \includegraphics[width=\textwidth]{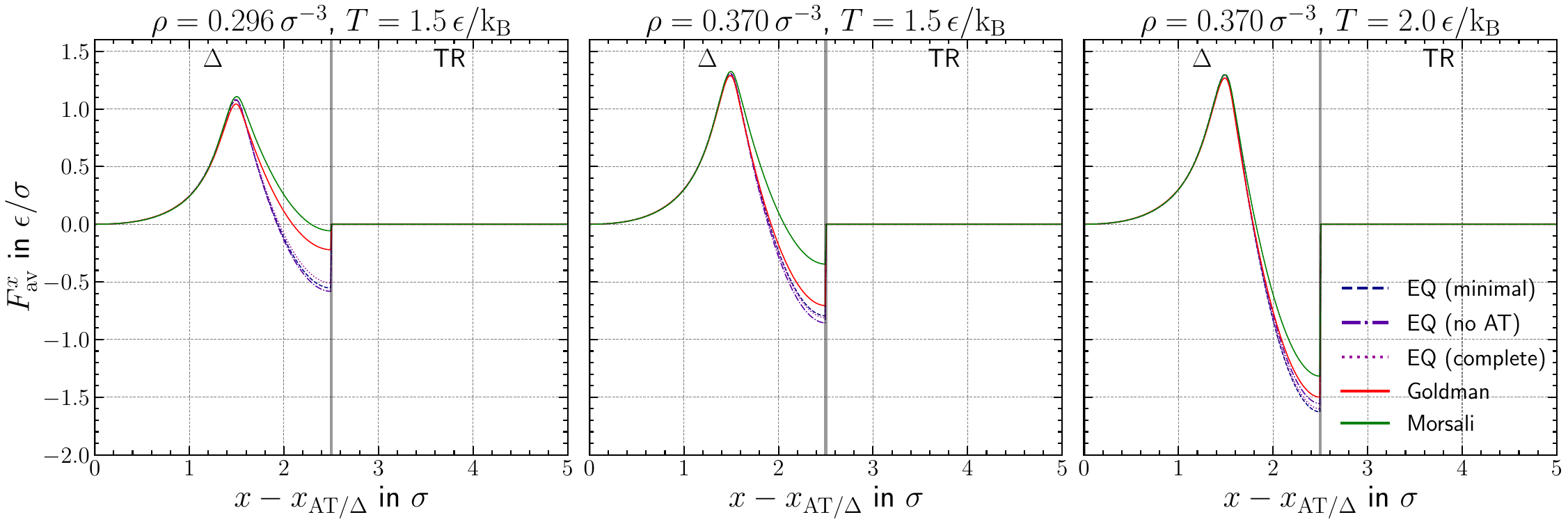}
    \caption{\label{fig:tf_apriori_array}Initial guesses for the thermodynamic force iterations. Each column represents one thermodynamic state. The forces were obtained as the average interface forces $F_{\textrm{av}}^x$ with the ``boundary'' placed at the position $x_{\textrm{$\Delta$/TR}}$ of the \textsc{$\Delta$/TR} interface marked with a vertical gray line. We used the radial distribution functions obtained in equilibration (\textsc{EQ}) runs in the minimal (dashed dark blue), no \textsc{AT} (dashed and dotted purple) or complete (dotted pink) box setups or taken from Morsali \textit{et al.} (green) or Goldman (red)\cite{morsali_accurate_2005, goldman_explicit_1979}. The $x$-axis was shifted by the position $x_{\textrm{AT/$\Delta$}}$ of the right-hand \textsc{AT/$\Delta$} boundary. Shown here are thus only the right-hand \textsc{$\Delta$} and \textsc{TR} regions, which are representative of the opposing left-hand $\Delta$ region due to symmetry.}
\end{figure}

In Fig. \ref{fig:tf_apriori_array} we compare the $F_{\textrm{av}}^x$'s calculated with the radial distribution functions obtained from the equilibrations of the three system setups that we introduced in section \ref{subsec:simulation_protocol} with the $F_{\textrm{av}}^x$'s calculated with the parametrized $g(r)$'s taken from Morsali \textit{et al.} and Goldman. It is clearly apparent that all the forces are in agreement with each other at a distance greater than $\SI{1}{\sigma}$ away from the \textsc{$\Delta$/TR} interface into the \textsc{$\Delta$} region. 

The region of interest for our hypothesis in Eq. (\ref{eq:F_th_close_to_boundary}) is anyways close to the \textsc{AT/$\Delta$} boundary because the part of the \textsc{$\Delta$} region close to the \textsc{$\Delta$/TR} interface where the $F_{\textrm{av}}^x$'s obtained from different $g(r)$'s deviate will be heavily influenced by the other artifacts mentioned in section \ref{sec:interface_corrections}. We conclude that we may use either one of these $F_{\textrm{av}}^x$'s as initial guess for the iteration procedure and the considerations to come. Consequently, we choose here the one obtained from the parametrized $g(r)$ of Goldman.

\subsection{Thermodynamic force iterations}\label{subsec:thermodynamic_force_iterations}
Using the \emph{a priori} approximation to the thermodynamic force as initial guess, we performed the iterative procedure of Eq. (\ref{eq:thermo_force_iterations}). At each iteration we calculated the density profile along the $x$-direction on an equidistant bin grid as described in section \ref{sec:interface_corrections}. 

The thermodynamic force was obtained from the negative numerical gradient of that density profile weighted by a convergence prefactor of $c = \SI{2}{\epsilon}/\rho$ (see Eq. (\ref{eq:thermo_force_iterations})). The resolution for the initial guesses was adapted to the bin grid. During the simulations, the thermodynamic force was applied to every particle within the \textsc{$\Delta$} and near \textsc{TR} regions (see app. \ref{app:simulation_details} for details) by linearly interpolating the force values given on the bin grid according to the respective particle position. 

We stopped the procedure once the density profile along the $x$-direction was sufficiently flat within the \textsc{$\Delta$} and \textsc{TR} regions and acquired the respective equilibrium value at the \textsc{AT/$\Delta$} boundary, which for all thermodynamic states and system setups was the $7$th iteration. 

We found that the behavior of both the converged thermodynamic force and the final density profile close to the \textsc{$\Delta$/TR} interface was very sensitive to the density bin width, but that the behavior of those two functions close to the \textsc{AT/$\Delta$} boundary was largely independent of the density bin width. 

\begin{figure}[ht]
    \centering
    \includegraphics[width=\textwidth]{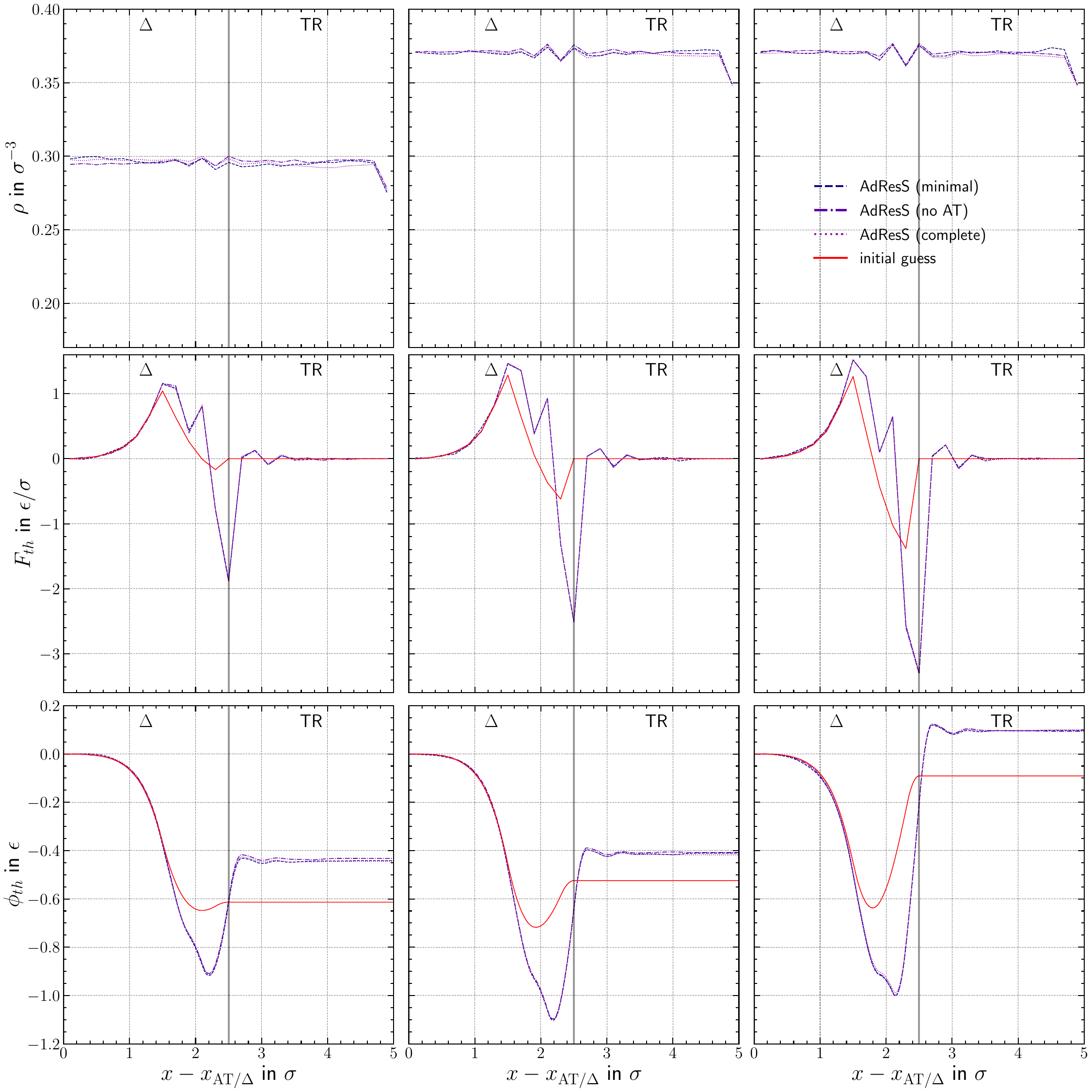}
    \caption{\label{fig:dens_tf_array}Results of the $7$th thermodynamic force iterations at the right-hand \textsc{$\Delta$/TR} interface, which is marked with a vertical gray line. Each column represents one thermodynamic state. Shown are the density profiles (first row), the thermodynamic forces (second row) and the corresponding coupling potentials (third row) for the minimal (dashed dark blue), no \textsc{AT} (dashed and dotted purple) and complete (dotted pink) setups and the initial guess obtained from Goldman's radial distribution function (red).}
\end{figure}

The density profiles $\rho(x)$, thermodynamic forces $F_{\textrm{th}}$ and corresponding coupling potentials $\phi_{\textrm{th}}$ obtained for the different thermodynamic states and system setups are shown in Fig. \ref{fig:dens_tf_array}. 

It is evident from the first row of Fig. \ref{fig:dens_tf_array} that for all states and system setups the density profiles are flat and at the equilibrium value throughout the entire \textsc{$\Delta$} and \textsc{TR} regions except very close to the \textsc{$\Delta$/TR} interface. We found that this oscillatory behavior could not be suppressed by running more iterations nor by changing the resolution at which the density profile is calculated. Such oscillations in the density profile have been observed in similar systems with discontinuous interfaces\cite{werder_hybrid_2005, kotsalis_control_2007, hansen_theory_2013}. 

In the second row of Fig. \ref{fig:dens_tf_array} we see the thermodynamic forces calculated from the iterative procedure compared with the respective initial guesses obtained from Goldman's $g(r)$'s as explained in section \ref{subsec:initial_guess}. It is here that we find confirmation that the proposed simplifications of section \ref{subsec:implications_for_the_thermodynamic_force} were justified: For every investigated thermodynamic state, the respective thermodynamic forces resulting from the iterative procedure for the minimal, no \textsc{AT} and complete box setups are practically indistinguishable from each other, implying that there is indeed no dependence of the thermodynamic force on the box dimensions $L_{y, z}$ or the width of the \textsc{AT} region $L_{\textrm{AT}}$. This appears to be valid even far away from the \textsc{AT/$\Delta$} boundary, indicating that also the effects of the excluded volume overlap and thermodynamic differences between \textsc{AT} and \textsc{TR} are independent of $L_{y, z}$ and $L_{\textrm{AT}}$. 

We did observe small deviations in some of the functions other than the thermodynamic force for simulations with even smaller box dimensions $L_{y, z} = \SI{10.0}{\sigma}$, which is in accordance with our statement that the independence of $F_{\textrm{th}}$ and $L_{y, z}$ only holds until finite-size effects become dominant. Note that such finite-size effects are not necessarily linear in the ratio of the relevant correlation length and the box size\cite{reible_finite-size_2023, reible_finite-size_2025, reible_fast_2025}. Even more so, the relevant correlation length is likely not the equilibrium correlation length of the Lennard-Jones fluid due to the non-trivial effects in between entering tracer particles. A detailed study of the finite-size effects is thus deferred to future work.   

Furthermore, we also find that the assumption of Eq. (\ref{eq:F_th_close_to_boundary}) and the validity of the initial guess are confirmed in the simulations: In the \textsc{$\Delta$} region close to the \textsc{AT/$\Delta$} boundary, the thermodynamic forces are indeed in very good agreement with the \emph{a priori} calculated initial guess, which we know from section \ref{subsec:initial_guess} to be a good approximation to $F_{\textrm{av}}^x$ of Eq. (\ref{eq:F_av_x}). Significant deviations only arise in the part of the \textsc{$\Delta$} region closer than $\SI{1}{\sigma}$ to the \textsc{$\Delta$/TR} interface and the \textsc{TR} region, where we do expect the interface artifacts other than the missing average force to have a substantial impact on the particles. We note that the thermodynamic forces within that region are strongly oscillatory, matching the behavior of the density profiles and that these oscillations appear remarkably similar across the different thermodynamic states, albeit with different amplitudes. 

We observed that, compared to the thermodynamic force iterations carried out without initial guess, the initial guess spared only one iteration and converged to the same result. If applied as is, the initial guess thus only marginally impacts the speed of convergence of the iterative procedure. However, considering that the procedure converged after only $7$ iterations, one iteration amounts to a substantial fraction of the overall cost.

We also observed that different widths $L_{\textrm{TR}}$ of the \textsc{TR} region impacted the behavior of the thermodynamic force close to the \textsc{$\Delta$/TR} interface, but not at the \textsc{AT/$\Delta$} boundary, which agrees well with our theoretical prediction of section \ref{subsec:implications_for_the_thermodynamic_force}. The size and nature of the \textsc{TR} region may, however impact finite size effects\cite{heidari_fluctuations_2018, heidari_finite-size_2018}. This may thus serve as motivation for future investigations.

In the third row of Fig. \ref{fig:dens_tf_array} we show the coupling potentials obtained by numerically integrating the thermodynamic forces. It is evident that there is a difference in coupling potential at the \textsc{AT/$\Delta$} boundary and in the deep \textsc{TR} region and that this difference has a strong dependence on the density and temperature. This is to be expected based on the fact that the Lennard-Jones fluid responds differently to the excluded volume overlap effect at the \textsc{$\Delta$/TR} interface and changes in the thermodynamic state than the tracer fluid found within the deep \textsc{TR} region.

\subsection{Validation of the calculated thermodynamic forces}
In section \ref{subsec:implications_for_the_thermodynamic_force}, we proposed a new set of observables to test the validity of the thermodynamic force that are adjusted to the interfacial nature of the AdResS method and do not require the presence of an \textsc{AT} region. 
The first of the criteria, namely the density profile, we already checked during the calculation stage of the thermodynamic force and we displayed the respective profiles in Fig. \ref{fig:dens_tf_array}. 

\begin{figure}[ht]
    \centering
    \includegraphics[width=\textwidth]{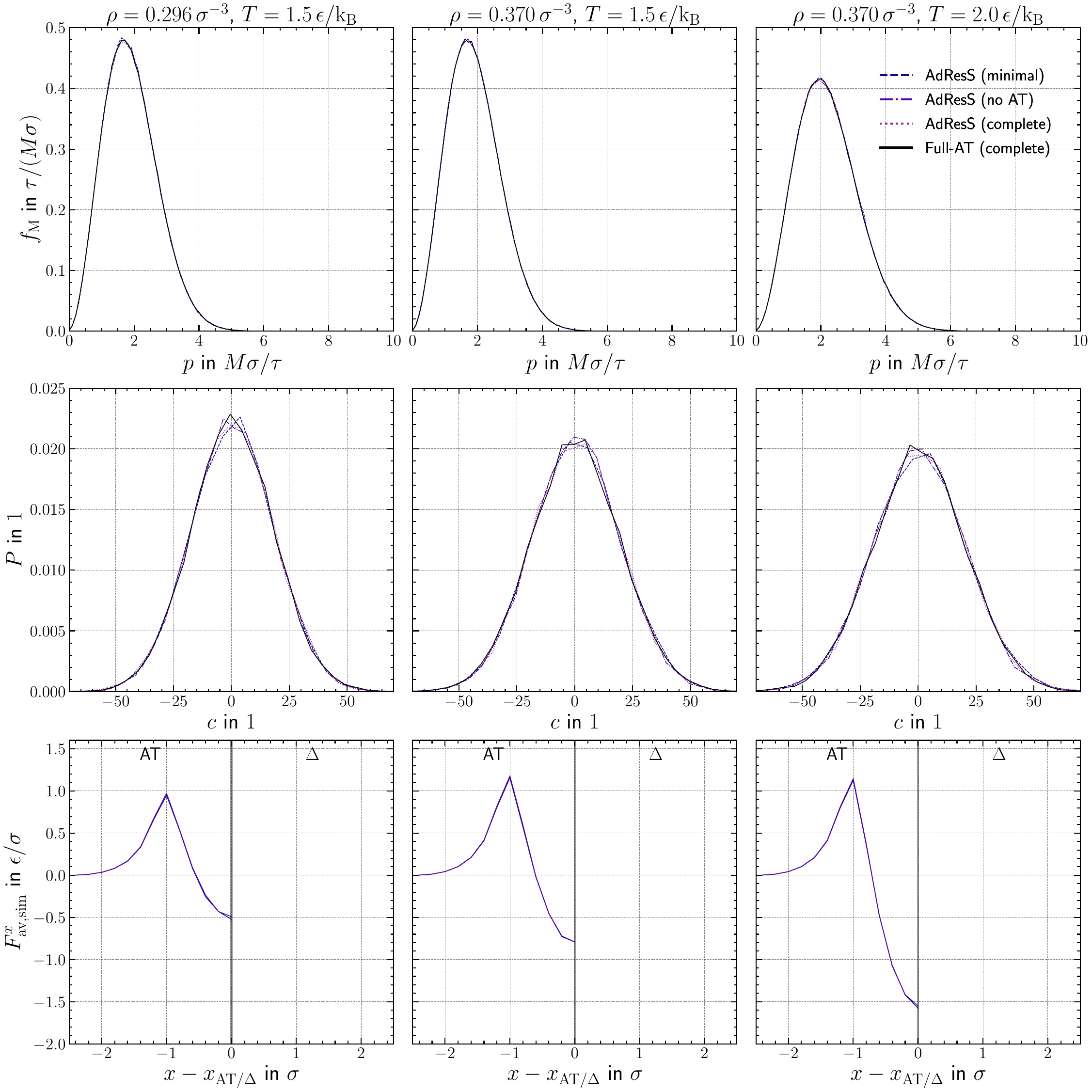}
    \caption{\label{fig:int_obs_array}Results of the validation runs at the right-hand \textsc{AT/$\Delta$} boundary. Each column represents one thermodynamic state. Shown are the momentum distributions (first row), the rescaled net numbers of particle crossings through the \textsc{AT/$\Delta$} boundary (second row) and the measured average force through the \textsc{AT/$\Delta$} boundary (third row) for the minimal (dashed dark blue), no \textsc{AT} (dashed and dotted purple) and complete (dotted pink) setups and the reference full-atomistic simulation (black).}
\end{figure}

Beyond that, we performed extended simulations using the iteratively calculated thermodynamic forces of the last section, measuring the observables in regular snapshots. We show the results in Fig. \ref{fig:int_obs_array}.

In the first row of Fig. \ref{fig:int_obs_array}, we see that for either thermodynamic state, the Maxwell-Boltzmann momentum distributions are almost indistinguishable across the different box setups, including the full-atomistic reference simulations. This confirms that the particles indeed cross the interface with the correct momentum distribution and, together with the density at the interface, adhere to the correct one-body statistics as required from the modeling assumption of the Liouville-type hierarchy in Eq. (\ref{eq:f_res_decorrelated}). 

The second row shows the distributions of the net number of particle crossings over long times rescaled for comparison between different box sizes as explained in appendix \ref{app:net_flux}. All distributions have zero mean, are of Gaussian shape and closely resemble the respective full-atomistic reference distribution. Emphasis is put on the observation that the smaller interface of the minimal setup also displays the same distribution (modulo the scaling). From this we can conclude that the exchange of particles through the interface exhibits the correct statistics up to second order, that is, mean value and fluctuations.

The last validating observable was the average force through the \textsc{AT/$\Delta$} boundary as required from Eq. (\ref{eq:avForce}) and it is shown in the third row of Fig. \ref{fig:int_obs_array}. We can again confirm that across all box setups, including the full-atomistic reference, these average forces are in very good agreement with each other for all three thermodynamic states. In particular, this means that the particles within the \textsc{$\Delta$} regions are indeed already sufficiently distributed in space to exert this correct average force through the \textsc{AT/$\Delta$} boundary onto the particles within \textsc{AT} - or even onto the particles in the adjacent \textsc{$\Delta$} region for the no \textsc{AT} and minimal setups.

We conclude that the respective thermodynamic force, together with the employed Langevin thermostat and the force capping, is capable of compensating the  artifacts of the \textsc{$\Delta$/TR} interface sufficiently fast such that, close to the \textsc{AT/$\Delta$} boundary, the only artifact left to compensate is the missing interactions as we claimed in Eq. (\ref{eq:F_th_close_to_boundary}). This is then accomplished by reintroducing the average interactions that particles in the \textsc{$\Delta$} region would be subject to through the \textsc{$\Delta$/TR} interface if the interactions with the particles in the \textsc{TR} region had not been switched off. We then find that the presence of the \textsc{$\Delta$/TR} interface does not affect the measured particle statistics at the \textsc{AT/$\Delta$} boundary and that the \textsc{AT/$\Delta$} boundary is thus a physically consistent boundary of an open system in accordance with the modeling of the reservoir given by Eqs. (\ref{eq:psi}) and (\ref{eq:phi}) of the Liouville-type hierarchy.

It should be noted that the thermal equilibrium that we assumed in our derivations is not, as in the Liouville-type hierarchy, enforced purely by the reservoir statistics, but rather by a globally applied Langevin thermostat. This may have implications on the dynamics of the particles and on the particle crossing statistics at the \textsc{AT/$\Delta$} boundary. It would thus be preferable to apply thermostatting without disturbing the particle dynamics at the boundary or within the \textsc{AT} region. Other studies have used local Langevin thermostats to achieve that\cite{ebrahimi_viand_theory_2020, klein_nonequilibrium_2021}. Our validation is, however, still consistent, as we applied the global thermostat to our full-atomistic reference as well. 

Another point of concern is the following: Even though we could confirm that the proposed simplifications for the thermodynamic force iterations led to the expected reduction in computational effort (see Tab. \ref{tab:simulation_runtimes}), the runtimes of the full-atomistic simulations were generally shorter than those of the respective AdResS simulations in the complete setup. This may be attributed to (1) the fact that the thermodynamic force needs to be applied as an additional one-body force, which introduces an additional scaling with the number of particles that counteracts the possible performance gains of switching off the interactions in the TR region; (2) the fact that the pair-wise force in AdResS is a function not only of the distance in between the particles, but also of the positions of both these particles, which introduces additional operations in the force calculation routines and may also impact the possible gains of the switching; (3) the excluded volume overlap at the \textsc{$\Delta$/TR} interfaces, due to which the particles there can acquire large velocities, in turn moving wider distances and increasing the frequency of the neighbor lists being rebuilt, which introduces computational cost that would not be present in full-atomistic simulations. As in this study we put the emphasis on the conceptual similarities between the Liouville-type hierarchy and the AdResS method and proposed simplifications that are independent of the implementation of AdResS, we defer assessing computational performance of the AdResS implementation to future work.  
\section{Conclusions}
We have shown that the behavior of the thermodynamic force close to the \textsc{AT/$\Delta$} boundary is well-captured by the missing average force $F_{\mathrm{av}}^x(x_{\Delta/\mathrm{TR}} - x)$ through the \textsc{$\Delta$/TR} interface. We have further provided a first-principles derivation of this average force starting from the Liouville-type hierarchy for the open system, arriving at a form that we could analyze for dependencies on important simulation parameters. This inspired a set of propositions to simplify the iterative procedure employed to calculate the thermodynamic force before any AdResS production run and new validation criteria for the thermodynamic force that are suited to the interfacial nature of the problem. 

\begin{table}[ht]
    \caption{\label{tab:dependencies_close_to_ATDelta}Dependencies of the thermodynamic force close to the \textsc{AT/$\Delta$} boundary.}
    \setlength{\arrayrulewidth}{0.2mm}
    \setlength{\tabcolsep}{6pt}
    \renewcommand{\arraystretch}{1.5}
    \centering
    \begin{tabular}{ l  c  c }
        \hline
        \textbf{Property} & \textbf{Dependence expected?} & \textbf{Coverage by our simulations} \\
        \hline
        Density $\rho$ &                                yes &           $\checkmark$ \\
        Temperature $T$ &                               yes &           $\checkmark$ \\
        Potential $V$   &                               yes &           $-$ \\
        Cut-off radius $r_{\mathrm{cut}}$ &             yes &           $-$ \\
        Capping radius $r_{\mathrm{cap}}$ &             no &            $-$ \\
        Box dimensions $L_y$ and $L_z$ &                no &            $\checkmark$ \\
        \textsc{AT} region width $L_{\textrm{AT}}$ &    no &            $\checkmark$ \\
        $\Delta$ region width $L_{\Delta}$ &            no &            $-$ \\
        Thermostat coupling parameter $\gamma$ &        no &            $-$ \\
        \textsc{TR} region properties &                          no &            $-$ \\
        \hline
    \end{tabular}
\end{table}

In a simulation study, we confirmed the aforementioned propositions and validation criteria for an array of different thermodynamic states and system setups. To summarize our results, we show an overview over the expected dependencies of the thermodynamic force close to the \textsc{AT/$\Delta$} boundary and the coverage of our simulations in Tab. \ref{tab:dependencies_close_to_ATDelta}. In particular, we have shown that a minimal AdResS box that does not contain an \textsc{AT} region altogether is sufficient to calculate the thermodynamic force and to simulate a physically consistent \textsc{AT/$\Delta$} boundary.

This work provides a close connection between the Liouville-type hierarchy for the open system and the AdResS method, reassuring both the validity of the modeling assumptions that went into the former and the physical soundness of the latter. It may also serve to inspire further simplifications for the simulation setup based on the parameters of Tab. \ref{tab:dependencies_close_to_ATDelta} that we did not cover in our survey. 

All our results were obtained assuming a single-species simple atomistic fluid such as the supercritical Lennard-Jones fluid. AdResS has been applied to a wide range of soft-matter systems with much more complex interactions and molecular structures or with multiple species present in the simulation (see ref. \cite{krekeler_adaptive_2018} and references therein). It will thus be an interesting next step to apply the formalism developed here to such systems, e.g. water, the tetrahedral Lennard-Jones molecular fluid or a multi-species Lennard-Jones fluid. Such applications could also serve as motivation to further expand the modeling assumptions of the Liouville-type hierarchy, which could be a timely contribution to the field of classical open systems\cite{winkelmann_stochastic_2020, del_razo_probabilistic_2022, del_razo_dynamics_2025}.

\begin{acknowledgments}
This research has been funded by Deutsche Forschungsgemeinschaft (DFG) through grant CRC 1114 Scaling Cascades in Complex Systems, Project Number 235221301, Project C01 Adaptive coupling of scales in molecular dynamics and beyond to fluid dynamics.

We acknowledge access to high-performance computers via the Zentraleinrichtung für Datenverarbeitung (ZEDAT) of Freie Universität Berlin.

We express our dearest gratitude towards Sebastian Eibl for helping us implement our simulations in his software package \textsc{MRMD} and towards Felix Höfling, Tetsuya Yamamoto, Tom Dörffel and the groups of Rupert Klein, Luigi Delle Site and Robinson Cortes-Huerto for insightful discussions.
\end{acknowledgments}
\bibliography{references.bib} 

\appendix
\section{Simulation details}\label{app:simulation_details}
We performed molecular dynamics simulations of a fluid with truncated, shifted Lennard-Jones potential in reduced units ($\sigma = \epsilon = m = 1)$. This potential is well-established in the modeling of Argon\cite{allen_computer_2017, frenkel_understanding_2023} and comparison with other simulations for Argon can be achieved by converting back to \textsc{SI} units.

We employed a cut-off radius of $r_{\textrm{cut}} = \SI{2.5}{\sigma}$ and a capping radius of $r_{\textrm{cap}} = \SI{0.82417464}{\sigma}$, which corresponds to capping the absolute pair-wise force at $F_{\textrm{cap}} = \SI{500}{\epsilon/\sigma}$. All simulations were done with stochastic dynamics integrator in the BAOAB scheme\cite{leimkuhler_molecular_2015} with a time step of $\mathrm{tstep} = \SI{0.002}{\tau}$ and Langevin thermostat applied everywhere in the box with coupling parameter $\mathrm{gamma} = \SI{20.0}{\tau^{-1}}$ implemented in the c++ software package \textsc{MRMD}, which is available on GitHub\cite{eibl_multi_2025}. The exact scripts that were used for the simulations can found at the in-house fork of MRMD (see \url{https://github.com/J-Hizzle/mrmd/tree/39e4117dec7ed36e97e99ffabe6c506d5925c3ae/examples/04_LennardJones_GentleChemostat}). The unit of time is given as $\tau = \sigma \sqrt{\frac{m}{\epsilon}}$.

\begin{table}[ht]
    \caption{\label{tab:simulation_parameters_general}General simulation parameters.}
    \setlength{\arrayrulewidth}{0.2mm}
    \setlength{\tabcolsep}{6pt}
    \renewcommand{\arraystretch}{1.5}
    \centering
    \begin{tabular}{ l  r }
        \hline
        \textbf{Parameter} & \textbf{Value} \\
        \hline
        $\sigma$ & $1$ \\
        $\epsilon$ & $1$ \\
        $m$ & $1$ \\
        $r_{\textrm{cut}}$ & $\SI{2.5}{\sigma}$ \\
        $r_{\textrm{cap}}$ & $\SI{0.82417464}{\sigma}$ \\
        $\mathrm{tstep}$ & $\SI{0.002}{\tau}$ \\
        $\mathrm{gamma}$ & $\SI{20.0}{\tau^{-1}}$ \\
        \hline
    \end{tabular}
\end{table}

We investigated three different AdResS box setups, namely (1: minimal) $L_{\textrm{AT}} = \SI{0.0}{\sigma}$, $L_x = \SI{10.0}{\sigma}$, $L_{y, z} = \SI{20.0}{\sigma}$, (2: no \textsc{AT}) $L_{\textrm{AT}} = \SI{0.0}{\sigma}$, $L_x = \SI{10.0}{\sigma}$, $L_{y, z} = \SI{30.0}{\sigma}$ and (3: complete) $L_{\textrm{AT}} = \SI{20.0}{\sigma}$, $L_x = \SI{30.0}{\sigma}$, $L_{y, z} = \SI{30.0}{\sigma}$, each for three different thermodynamic states with (1: low density, low temperature) $\rho = \SI{0.296}{\sigma^3}$ and $T = \SI{1.5}{\epsilon/k_{\textrm{B}}}$, (2: high density, low temperature) $\rho = \SI{0.370}{\sigma^3}$ and $T = \SI{1.5}{\epsilon/k_{\textrm{B}}}$ and (3: high density, high temperature) $\rho = \SI{0.370}{\sigma^3}$ and $T = \SI{2.0}{\epsilon/k_{\textrm{B}}}$.

\begin{table}[ht]
    \caption{\label{tab:thermodynamic_states}Thermodynamic states.}
    \setlength{\arrayrulewidth}{0.2mm}
    \setlength{\tabcolsep}{6pt}
    \renewcommand{\arraystretch}{1.5}
    \centering
    \begin{tabular}{ l  r  r  r }
        \hline
        \textbf{Property} & \textbf{low $\rho$, low $T$} & \textbf{high $\rho$, low $T$} & \textbf{high $\rho$, high $T$}\\
        \hline
        Density $\rho$ & $\SI{0.296}{\sigma^{-3}}$ & $\SI{0.370}{\sigma^{-3}}$ & $\SI{0.370}{\sigma^{-3}}$ \\
        Temperature $T$ & $\SI{1.5}{\epsilon/k_{\textrm{B}}}$ & $\SI{1.5}{\epsilon/k_{\textrm{B}}}$ & $\SI{2.0}{\epsilon/k_{\textrm{B}}}$ \\
        \hline
    \end{tabular}
\end{table}

The respective total numbers of particles for each state and box setup were (1: minimal box at low density and low temperature) $N = 1184$, (2: no \textsc{AT} box at low density and low temperature) $N = 2664$, (3: complete box at low density and low temperature) $N = 7992$, (4: minimal box at high density and low temperature) $N = 1480$, (5: no \textsc{AT} box at high density and low temperature) $N = 3330$, (6: complete box at high density and low temperature) $N = 9990$, (7: minimal box at high density and high temperature) $N = 1480$, (8: no \textsc{AT} box at high density and high temperature) $N = 3330$, (9: complete box at high density and high temperature) $N = 9990$.

\begin{table}[b]
    \caption{\label{tab:simulation_parameters_box_setup}Parameters for the box setups and thermodynamic states.}
    \setlength{\arrayrulewidth}{0.2mm}
    \setlength{\tabcolsep}{6pt}
    \renewcommand{\arraystretch}{1.5}
    \centering
    \begin{tabular}{ l  r  r  r }
        \hline
        \textbf{Parameter} & \textbf{AdResS (minimal)} & \textbf{AdResS (no \textsc{AT})} & \textbf{AdResS (complete)} \\
        \hline
        $L_{\textrm{AT}} $ & $\SI{0.0}{\sigma}$ & $\SI{0.0}{\sigma}$ & $\SI{20.0}{\sigma}$ \\
        $L_x$ & $\SI{10.0}{\sigma}$ & $\SI{10.0}{\sigma}$ & $\SI{30.0}{\sigma}$ \\
        $L_{y, z}$ & $\SI{20.0}{\sigma}$ & $\SI{30.0}{\sigma}$ & $\SI{30.0}{\sigma}$ \\
        $L_{\textrm{$\Delta$}}$ & $\SI{2.5}{\sigma}$ & $\SI{2.5}{\sigma}$ & $\SI{2.5}{\sigma}$ \\
        $L_{\textrm{TR}}$ & $\SI{5.0}{\sigma}$ & $\SI{5.0}{\sigma}$ & $\SI{5.0}{\sigma}$ \\ 
        $x_{\textrm{center}}$ & $\SI{5.0}{\sigma}$ & $\SI{5.0}{\sigma}$ & $\SI{15.0}{\sigma}$ \\
        $r_{\textrm{AT/$\Delta$}}$ & $\SI{0.0}{\sigma}$ & $\SI{0.0}{\sigma}$ & $\SI{10.0}{\sigma}$ \\
        $r_{\textrm{$\Delta$/TR}}$ & $\SI{2.5}{\sigma}$ & $\SI{2.5}{\sigma}$ & $\SI{12.5}{\sigma}$ \\
        $N$ ($\rho = \SI{0.296}{\sigma^3}$, $T = \SI{1.5}{\epsilon/k_B}$) & $1184$ & $2664$ & $7992$ \\
        $N$ ($\rho = \SI{0.370}{\sigma^3}$, $T = \SI{1.5}{\epsilon/k_B}$) & $1480$ & $3330$ & $9990$ \\
        $N$ ($\rho = \SI{0.370}{\sigma^3}$, $T = \SI{2.0}{\epsilon/k_B}$) & $1480$ & $3330$ & $9990$ \\
        \hline
    \end{tabular}
\end{table}

We calculated the \textit{a priori} approximations shown in Fig. \ref{fig:tf_apriori_array} by evaluating the integral in Eq. (\ref{eq:F_av_x}) with Gaussian quadrature, using $x_{\textrm{$\Delta$/TR}}$ for the position of the boundary and inserting radial distribution functions either (1) obtained in the equilibration step of each respective box setup and interpolated with cubic B-Splines (\textsc{EQ}: minimal, no \textsc{AT} and complete) or (2) taken in parametrized form from the publications of Morsali \textit{et. al} and Goldman (Morsali and Goldman)\cite{morsali_accurate_2005, goldman_explicit_1979}. The evaluation and processing was done with the in-house python package GhostFace, which is publicly available on GitHub (see \url{https://github.com/J-Hizzle/GhostFace/blob/e9dbc30007614809880fba4ede1f893a1a49f33f/examples/calc_initial_guess.py}).

The thermodynamic forces were then calculated in all thermodynamic states and box setups using the iterative procedure of Eq. (\ref{eq:thermo_force_iterations}). We performed $\mathrm{niter} = 10$ iterations for each thermodynamic force, where each iteration comprised an AdResS simulation over $\mathrm{update} = 5000000$ time steps. During each iteration, the density profile was sampled in slabs of width $\mathrm{densbinwidth} = \SI{0.2}{\sigma}$ every $\mathrm{sampling} = 500$ simulation steps and then averaged. The force value at each density grid point was calculated by (1) symmetrizing the averaged density profile by taking the average of the left- and right-hand values with equal distance from the center of the box, (2) calculating a numerical gradient of the symmetrized density profile in the form of a symmetric difference quotient, (3) setting the gradient values to zero outside and at the boundaries of the application region of the force, (4) subtracting the gradient multiplied with $\mathrm{forcemod} = \SI{2}{\epsilon}/\rho$ from the current thermodynamic force (this corresponds to setting the convergence prefactor $c$ of Eq. (\ref{eq:thermo_force_iterations}) to $\SI{2}{\epsilon}/\rho$). The application region was defined as the region starting at a distance $\mathrm{appmin} = r_{\textrm{AT/$\Delta$}}$ with $r_{\textrm{AT/$\Delta$}} = x_{\textrm{AT/$\Delta$}} - x_{\textrm{center}}$ from the box center $x_{\textrm{center}}$ in $x$-direction to $\mathrm{appmax} = r_{\textrm{AT/$\Delta$}} + \SI{4.5}{\sigma}$. The \textsc{$\Delta$} and \textsc{TR} regions were used with a width of $L_{\Delta} = \SI{2.5}{\sigma}$ and $L_{\textrm{TR}} = \SI{5.0}{\sigma}$, respectively. The initial guesses were obtained from Goldman's radial distribution functions, evaluated on the density grid. 

The respective thermodynamic force was applied at every timestep to every particle currently residing within the application region by linearly interpolating the values given on the grid according to the position of that particle.

All iterations were considered converged after the $7$th iteration and the obtained thermodynamic forces were used for the subsequent validation runs accordingly. 

The postprocessing and plotting of the \textsc{MRMD} output data was performed by the in-house python package MRMDAnalysis, which is publicly available on GitHub (see \url{https://github.com/J-Hizzle/MRMDAnalysis/tree/a323524131f3a4870a12b79de476f9d0d4517aef/examples}). The densities, thermodynamic forces and corresponding coupling potentials of the converged $7$th iterations are shown in Fig. \ref{fig:dens_tf_array}.  

\begin{table}[t]
    \caption{\label{tab:simulation_parameters_iterations}Simulation parameters for thermodynamic force iterations. The thermodynamic force obtained from the $7$th iteration was considered converged and used in the subsequent validation runs.}
    \setlength{\arrayrulewidth}{0.2mm}
    \setlength{\tabcolsep}{6pt}
    \renewcommand{\arraystretch}{1.5}
    \centering
    \begin{tabular}{ l  r }
        \hline
        \textbf{Parameter} & \textbf{Value} \\
        \hline
        $\mathrm{niter}$ & $7$/$10$ \\
        $\mathrm{update}$ & $5000000$ \\
        $\mathrm{sampling}$ & $500$ \\
        $\mathrm{densbinwidth}$ & $\SI{0.2}{\sigma}$ \\
        $\mathrm{appmin}$ & $r_{\textrm{AT/$\Delta$}}$ \\
        $\mathrm{appmax}$ & $r_{\textrm{AT/$\Delta$}} + 4.5 \sigma$ \\
        $\mathrm{forcemod}$ & $\SI{2}{\epsilon}/\rho$ \\
        \hline
    \end{tabular}
\end{table}

The validation simulations were performed in each state and setup using the thermodynamic force obtained in the respective $7$th iteration step and running over $\mathrm{nsteps} = 200000001$ time steps. At each step, the net number of particles crossing the right-hand \textsc{AT/$\Delta$} boundary was counted (a particle moving through the interface from left to right (out of the open system) was counted as $-1$ and from right to left (into the open system) as $+1$ and added to the numbers counted at the preceding steps. Every $\mathrm{output} = 10000$ steps, the sum over net numbers of particle crossings was stored (along with the positions, velocities and forces of the particles) and then reset to $0$. These datasets were then processed and plotted with MRMDAnalysis and can be seen in Fig. \ref{fig:int_obs_array}.

\begin{table}[b]
    \caption{\label{tab:simulation_parameters_validation}Simulation parameters for validation runs.}
    \setlength{\arrayrulewidth}{0.2mm}
    \setlength{\tabcolsep}{6pt}
    \renewcommand{\arraystretch}{1.5}
    \centering
    \begin{tabular}{ l  r }
        \hline
        \textbf{Parameter} & \textbf{Value} \\
        \hline
        $\mathrm{nsteps}$ & $200000001$ \\
        $\mathrm{output}$ & $10000$ \\
        \hline
    \end{tabular}
\end{table}

We list the total runtimes for each simulation in tab. \ref{tab:simulation_runtimes}.
\begin{table}[ht]
    \caption{\label{tab:simulation_runtimes}Total simulation runtimes on a single-core GPU node with Nvidia GeForce 1080 on the Curta cluster \cite{bennett_curta_2020}. For the iterations, the numbers before the slash are the times at which the $7$th iteration had finished, which was then used in the subsequent validation runs.}
    \setlength{\arrayrulewidth}{0.2mm}
    \setlength{\tabcolsep}{6pt}
    \renewcommand{\arraystretch}{1.5}
    \centering
    \begin{tabular}{ l | l  r  r  r  r }
        \hline
        \textbf{Simulation} & \textbf{State} & \textbf{AdResS (minimal)} & \textbf{AdResS (no \textsc{AT})} & \textbf{AdResS (complete)} & \textbf{FA (complete)} \\
        \hline
        \textbf{2} Iteration & low $\rho$, low $T$ & $2.76$/$\SI{3.94}{h}$ & $2.74$/$\SI{3.92}{h}$ & $3.47$/$\SI{4.95}{h}$ & - \\
                             & high $\rho$, low $T$ & $2.48$/$\SI{3.54}{h}$ & $2.66$/$\SI{3.81}{h}$ & $4.21$/$\SI{6.02}{h}$ & - \\
                             & high $\rho$, high $T$ & $2.72$/$\SI{3.88}{h}$ & $3.24$/$\SI{4.64}{h}$ & $4.10$/$\SI{5.86}{h}$ & - \\
        \textbf{3} Validation & low $\rho$, low $T$ & $\SI{14.38}{h}$ & $\SI{15.56}{h}$ & $\SI{22.07}{h}$ & $\SI{19.11}{h}$ \\
                              & high $\rho$, low $T$ & $\SI{15.05}{h}$ & $\SI{16.70}{h}$ & $\SI{23.11}{h}$ & $\SI{22.91}{h}$ \\
                              & high $\rho$, high $T$ & $\SI{15.06}{h}$ & $\SI{18.96}{h}$ & $\SI{27.32}{h}$ & $\SI{23.60}{h}$ \\
        \hline 
    \end{tabular}
\end{table}

\section{Measuring the distribution of the net number of particle crossings}\label{app:net_flux}
\subsection{Relations between normal distributions}
The Gaussian normal distribution with mean $\mu$ and standard deviation $\sigma$ is defined as (see Eq. (23.56) on p. 1155 of ref. \cite{arfken_mathematical_2012})
\begin{align}
    f(x; \mu, \sigma) &= \frac{1}{\sqrt{2 \pi \sigma^2}} \exp{-\frac{(x - \mu)^2}{2 \sigma^2}} \textrm{,}
\end{align}
which is related to the standard Gaussian normal distribution
\begin{align}
    \phi(x) &= \frac{1}{\sqrt{2 \pi}} \exp{-\frac{x^2}{2}}
\end{align}
by stretching the domain of $\phi$ by a factor of $\sigma$, subsequently shifting it by $\mu$ and normalizing $\phi$ by $1/\sigma$ according to
\begin{align}\label{eq:general_to_standard_gauss}
    f(x; \mu, \sigma) &= \frac{1}{\sigma} \phi\left(\frac{x - \mu}{\sigma}\right) \textrm{.}
\end{align}
This can be used to compare two normal distributions $f_1$ and $f_2$ with different standard deviations $\sigma_1$ and $\sigma_2$ and zero mean $\mu_1 = \mu_2 = 0$ with each other by making use of Eq. (\ref{eq:general_to_standard_gauss}) twice as
\begin{align}\label{eq:compare_general_gaussian_distributions}
    f_2(x; 0, \sigma_2) &= \frac{1}{\sigma_2} \phi\left(\frac{x}{\sigma_2}\right) = \frac{\sigma_1}{\sigma_2} f_1\left(\frac{\sigma_1}{\sigma_2} x; 0, \sigma_1\right) \textrm{,}
\end{align}
from which we know that the two distributions are the same up to a stretching of the domain by a factor of $\sigma_2/\sigma_1$ and normalization by a factor of $\sigma_1/\sigma_2$. 

\subsection{Normal distributions from measurements}
From the central limit theorem (see appendix A.1.4.1 on p. 400 of ref. \cite{reichl_modern_2016}), we know that the distribution $f_{X_N}(x)$ of the sample mean $X_N = \frac{1}{N} \sum_{i = 1}^{N} Y_i$ for $N$ independent measurements of the identically distributed (IID) random variable $Y_i$ with unknown probability distribution $f'_Y$ which is centered at $0$ and has finite moments becomes a Gaussian normal distribution in the limit 
\begin{align}
    \lim_{N \rightarrow \infty} f_{X_N}(x) &= f_X(x; 0, \sigma_X) = \frac{1}{\sqrt{2 \pi \sigma_X^2}} \exp{-\frac{x^2}{2 \sigma_X^2}}
\end{align}
with standard deviation
\begin{align}\label{eq:clt_standard_deviation}
    \sigma_X = \frac{\sigma_Y}{\sqrt{N}} \textrm{.}
\end{align}
Here, $\sigma_Y$ is the standard deviation of the unknown distribution $f'_Y$. 

Stretching the domain of $f(x; 0, \sigma_X)$ and normalizing according to Eq. (\ref{eq:compare_general_gaussian_distributions}) yields a normal distribution 
\begin{align}\label{eq:compare_clt_gaussian_to_underlying}
    f_Y(x; 0, \sigma_Y) &= \frac{\sigma_X}{\sigma_Y} f_X\left(\frac{\sigma_X}{\sigma_Y} x; 0, \sigma_X\right) \nonumber\\
    &= \frac{1}{\sqrt{N}} f_X\left(\frac{1}{\sqrt{N}} x; 0, \sigma_X\right) \nonumber\\
    &= \frac{1}{\sqrt{2 \pi \sigma_Y^2}} \exp{-\frac{x^2}{2 \sigma_Y^2}} \textrm{,}
\end{align}
which has the same mean and standard deviation as the unknown distribution $f'_Y$.

Rather than using the sample mean $X_N$, one may sum over the $N$ IID variables $Y_i$ as $X'_N = N X_N$, which yields a normal distribution (see chapter I.7 on p. 26 of ref. \cite{van_kampen_stochastic_2007})
\begin{align}
    \lim_{N \rightarrow \infty} f_{X'_N}(x) &= f_{X'}(x; 0, \sigma_{X'})
\end{align}
with standard deviation 
\begin{align}
    \sigma_{X'} &= \sqrt{N} \sigma_Y = N \sigma_X \textrm{.}
\end{align}

Following the reasoning of Eq. (\ref{eq:compare_general_gaussian_distributions}), this can be related to the distribution $f_X$ by  
\begin{align}\label{eq:compare_sum_normal_distribution}
    f_{X'}(x; 0, \sigma_{X'}) &= \frac{\sigma_X}{\sigma_{X'}} f_X\left(\frac{\sigma_X}{\sigma_{X'}} x; 0, \sigma_X\right) \nonumber\\
    &= \frac{1}{N} f_X\left(\frac{1}{N} x; 0, \sigma_X\right) \textrm{.}
\end{align}

If we have two distributions $f_{X}^N$ and $f_{X}^M$ with standard deviations $\sigma_{X}^N$ and $\sigma_{X}^M$, sample means $X_N$ and $X_M$ and with $N$ and $M$ sufficiently large such that they can both be approximated well by Gaussian normal distributions with the same underlying distribution $f'_Y$ in the sense of Eqs. (\ref{eq:clt_standard_deviation}) and (\ref{eq:compare_clt_gaussian_to_underlying}), it is clear from Eq. (\ref{eq:compare_general_gaussian_distributions}) that they can be compared via 
\begin{align}\label{eq:compare_general_clt_gaussians}
    f_{X}^M(x; 0, \sigma_{X}^M) &= \sqrt{\frac{M}{N}} f_{X}^N\left(\sqrt{\frac{M}{N}} x; 0, \sigma_{X}^N\right) \textrm{.}
\end{align}

Equivalently, if both distributions have been obtained with the sample sums $X'_N$ and $X'_M$, they can be compared via
\begin{align}\label{eq:eq:compare_general_sum_clt_gaussians}
    f_{X'}^M(x; 0, \sigma_{X'}^M) &= \sqrt{\frac{N}{M}} f_{X'}^N\left(\sqrt{\frac{N}{M}} x; 0, \sigma_{X'}^N\right) \textrm{.}
\end{align}

This can be rephrased as a measurement protocol: If one measures two distributions in an experiment that both adhere to the central limit theorem, rescales one of them according to Eq. (\ref{eq:eq:compare_general_sum_clt_gaussians}) and finds that they become the same normal distribution, one can conclude that the underlying statistics $f'_Y$ is the same for both up to second order, that is, in mean and fluctuations.   

\subsection{Distributions of particle crossings}
Let $C_{a, k}^{L, \Delta \tau}$ be the net number of particle crossings through a plane $a$ parallel to the $yz$-plane with area $A = L^2$ within a time interval $k$ with duration $\Delta \tau$ measured in an equilibrium molecular fluid. Each particle crossing from right to left is counted as $+1$ (entering the open system) and each particle crossing from left to right is counted as $-1$ (exiting the open system). It is clear that the equilibrium mean of $C_{a, k}^{L, \Delta \tau}$ is identically zero. However, due to spontaneous and local fluctuations in the fluid, $C_{a, k}^{L, \Delta \tau}$ will be distributed around its mean according to an unknown probability distribution $P^{L, \Delta \tau}(c^{L, \Delta \tau})$ with standard deviation $\sigma_{L, \Delta \tau}$.

If $L$ and $\Delta \tau$ are sufficiently large such that during the measurement, there have occurred many uncorrelated particle crossings, $C_{a, k}^{L, \Delta \tau}$ can be considered an IID random variable. The sample sum
\begin{align}
    C_{O, T} &= \sum_{a = 1}^{O}\sum_{k = 1}^{T} C_{a, k}^{L, \Delta \tau} 
\end{align}
over many such measurements, taken over a long time interval $\Delta T = T\Delta \tau$ with $T$ different time steps and a large area $B = O A$ with $O$ different planes, thus adheres to the central limit theorem and its distribution $P_{O, T}(c)$ becomes a Gaussian normal distribution in the limit 
\begin{align}
    \lim_{OT \rightarrow \infty} P_{O, T}(c) &= f_C(c; 0, \sigma_C) = \frac{1}{\sqrt{2 \pi \sigma_C^2}} \exp{-\frac{c^2}{2 \sigma_C^2}} \textrm{,}
\end{align}
with standard deviation 
\begin{align}
    \sigma_C = \sqrt{O T} \sigma_{L, \Delta \tau} \textrm{.}
\end{align}
Following the reasoning of Eq. (\ref{eq:compare_general_clt_gaussians}), two such distributions $f_{C}^1$ and $f_{C}^2$ of sample sums $C_{1, 2}$ measured at two different numbers of planes $O_{1, 2}$ and time steps $T_{1, 2}$ that are both large enough to approximately yield normal distributions with respective standard deviations $\sigma_{C, 1}$ and $\sigma_{C, 2}$ and were obtained from the same underlying statistics can be compared via 
\begin{align}
    f_{C}^2(c; 0, \sigma_{C, 2}) &= \sqrt{\frac{O_1 T_1}{O_2 T_2}} f_{C}^1\left(\sqrt{\frac{O_1 T_1}{O_2 T_2}} c; 0, \sigma_{C, 1}\right) \textrm{.}
\end{align}

Since the underlying statistics are the same for both $f_{C}^1$ and $f_{C}^2$, the scaling factor can be expressed as 
\begin{align}
    \sqrt{\frac{O_1 T_1}{O_2 T_2}} = \sqrt{\frac{B_1 \Delta T_1}{B_2 \Delta T_2}} \textrm{,}
\end{align}
where $B_{1, 2} = O_{1, 2} A$ and $\Delta T_{1, 2} = T_{1, 2} \Delta \tau$. 

In case the two distributions $f_{C}^1$ and $f_{C}^2$ were measured over the same time interval $\Delta T_1 = \Delta T_2 = \Delta T = T \Delta \tau$ and at two square planes with different areas $B_{1, 2} = L_{1, 2}^2 = O_{1, 2} L^2$, the scaling factor simplifies to the ratio of the plane edge lengths $L_1$ and $L_2$ as 
\begin{align}
    \sqrt{\frac{O_1 T_1}{O_2 T_2}} = \frac{L_1}{L_2} \textrm{.}
\end{align}

In this case, 
\begin{align}\label{eq:compare_measured_gaussians}
    f_{C}^2(c; 0, \sigma_{C, 2}) &= \frac{L_1}{L_2} f_{C}^1\left(\frac{L_1}{L_2} c; 0, \sigma_{C, 1}\right) \textrm{.}
\end{align}

This is, again, to be understood as a measurement protocol: If one measures the distributions of sample sums $C_{1, 2}$, finds that both approximately yield Gaussian normal distributions $f_{C}^1$ and $f_{C}^2$, rescales one according to Eq. (\ref{eq:compare_measured_gaussians}) and finds that they match with each other, it follows that they share the same underlying statistics up to second order, which means that the mean and fluctuations of the net particle crossings are in agreement between the compared measurements. 

We followed this protocol in the validation of our thermodynamic forces, comparing the net particle crossing distributions obtained in the minimal box setup with the full-atomistic reference simulation obtained with the box dimensions of the complete setup. The results are shown in the second row of Fig. \ref{fig:int_obs_array}.

\section{Measuring the average force through the boundary}\label{app:F_av_x}
The average force through the boundary was measured on the same grid with the same binning as the density profile. Within each bin with center at $x_{\textrm{bin}}$ and width $2 x_{\textrm{m}}$ and at each stored snapshot of the trajectory at time step $s$, the instantaneous average pair-wise force $\vector{F}^{s, i}_{\textrm{av, sim}}$ through the boundary at $x_{\textrm{AT/$\Delta$}}$ exerted on a particle $i$ residing in the bin and left-hand of the boundary was calculated by dividing the sum of all instantaneous pair-wise forces $\vector{F}^{s, i, k}_{\textrm{av, sim}}$ through the boundary by the number $N_k$ of interacting particles labeled $k$ right-hand of the boundary as 
\begin{align}
    \vector{F}^{s, i}_{\textrm{av, sim}}(x_{\textrm{bin}}; x_{\textrm{AT/$\Delta$}}, x_{\textrm{m}}) &= \frac{1}{N_k} \sum_{k = 1}^{N_k} \vector{F}^{s, i, k}_{\textrm{av, sim}}(x_{\textrm{bin}}; x_{\textrm{AT/$\Delta$}}, x_{\textrm{m}}) \textrm{.}
\end{align}
The instantaneous average force $\vector{F}_{\textrm{av, sim}}^s$ per particle within the bin was then obtained by summing the $\vector{F}_{\textrm{av}, i}$ over all particles residing in the bin and dividing by their number $N_i$ as
\begin{align}
    \vector{F}_{\textrm{av, sim}}^s(x_{\textrm{bin}}; x_{\textrm{AT/$\Delta$}}, x_{\textrm{m}}) &= \frac{1}{N_i} \sum_{i = 1}^{N_i} \vector{F}^{s, i}_{\textrm{av, sim}}(x_{\textrm{bin}}; x_{\textrm{AT/$\Delta$}}, x_{\textrm{m}}) \textrm{.}
\end{align}
The total average force $\vector{F}_{\textrm{av, sim}}$ could then be calculated by summing $\vector{F}_{\textrm{av, sim}}^s$ over all time steps and dividing by their number $N_s$ as 
\begin{align}
    \vector{F}_{\textrm{av, sim}}(x_{\textrm{bin}}; x_{\textrm{AT/$\Delta$}}, x_{\textrm{m}}) &= \frac{1}{N_s} \sum_{s = 1}^{N_s} \vector{F}_{\textrm{av, sim}}^s(x_{\textrm{bin}}; x_{\textrm{AT/$\Delta$}}, x_{\textrm{m}}) \textrm{.}
\end{align}
For a full-atomistic reference simulation with insignificant finite-size effects, this total average force $\vector{F}_{\textrm{av, Full-AT}}$ converges to the average force of Eq. (\ref{eq:avForce}) in the limit of infinitely many time steps and infinitely fine binning such that we may write
\begin{align}
    \lim_{x_{\textrm{m}} \rightarrow 0} \lim_{N_s \rightarrow \infty} \vector{F}_{\textrm{av, Full-AT}}(x_{\textrm{bin}}; x_{\textrm{AT/$\Delta$}}, x_{\textrm{m}}) &= \vector{F}_{\textrm{av, Full-AT}}(x = x_{\textrm{bin}}; x_{\textrm{AT/$\Delta$}}) = \vector{F}_{\textrm{av}}(x = x_{\textrm{bin}}; x_{\textrm{AT/$\Delta$}}) \textrm{.}
\end{align}
As a consequence of the planar geometry of the \textsc{AT/$\Delta$} boundary, we know that the $y$- and $z$-components of $\vector{F}_{\textrm{av, Full-AT}}$ are identically zero, which we could confirm in our measurements. The $x$-components are non-trivial and are shown for all thermodynamic states in the third row of Fig. \ref{fig:int_obs_array}.
\end{document}